\documentclass[fleqn,usenatbib]{mnras}

\usepackage{newtxtext,newtxmath}
\usepackage[T1]{fontenc}
\usepackage{ae,aecompl}
\usepackage{lineno}
\usepackage{graphicx}	% Including figure files
\usepackage{amsmath}	% Advanced maths commands
\usepackage{amssymb}	% Extra maths symbols

\DeclareMathOperator*{\argmin}{arg\,min}
\usepackage{esvect} 

\usepackage{eso-pic}% http://ctan.org/pkg/eso-pic

\AddToShipoutPictureBG*{%
  \AtPageUpperLeft{%
    \hspace{0.75\paperwidth}%
    \raisebox{-3.5\baselineskip}{%
      \makebox[0pt][l]{\textnormal{DES 2017-0309}}
}}}%

\AddToShipoutPictureBG*{%
  \AtPageUpperLeft{%
    \hspace{0.75\paperwidth}%
    \raisebox{-4.5\baselineskip}{%
      \makebox[0pt][l]{\textnormal{FERMILAB-PUB-18-001-PPD}}
}}}%
\title[DES SV Mass Maps with Gaussian and Sparsity Priors]{Improving Weak Lensing Mass Map Reconstructions using Gaussian and Sparsity Priors: Application to DES SV}

\author[N. Jeffrey et al.]{
\parbox{\textwidth}{
\Large{N. Jeffrey,$^{1}$\thanks{E-mail: niall.jeffrey.15@ucl.ac.uk}
F.~B.~Abdalla$^{1,2}$,
O.~Lahav$^{1}$,
F.~Lanusse$^{3}$,
J.-L.~Starck$^{4}$,
A.~Leonard$^{1}$,
D.~Kirk$^{1}$,
C.~Chang$^{5}$,
E.~Baxter$^{6}$,
T.~Kacprzak$^{7}$,
S.~Seitz$^{8,9}$,
V.~Vikram$^{10}$,
L.~Whiteway$^{1}$,
T.~M.~C.~Abbott$^{11}$,
S.~Allam$^{12}$,
S.~Avila$^{13,14}$,
E.~Bertin$^{15,16}$,
D.~Brooks$^{1}$,
A.~Carnero~Rosell$^{17,18}$,
M.~Carrasco~Kind$^{19,20}$,
J.~Carretero$^{21}$,
F.~J.~Castander$^{22,23}$,
M.~Crocce$^{22,23}$,
C.~E.~Cunha$^{24}$,
C.~B.~D'Andrea$^{6}$,
L.~N.~da Costa$^{17,18}$,
C.~Davis$^{24}$,
J.~De~Vicente$^{25}$,
S.~Desai$^{26}$,
P.~Doel$^{1}$,
T.~F.~Eifler$^{27,28}$,
A.~E.~Evrard$^{29,30}$,
B.~Flaugher$^{12}$,
P.~Fosalba$^{22,23}$,
J.~Frieman$^{12,5}$,
J.~Garc\'ia-Bellido$^{14}$,
D.~W.~Gerdes$^{29,30}$,
D.~Gruen$^{24,31}$,
R.~A.~Gruendl$^{19,20}$,
J.~Gschwend$^{17,18}$,
G.~Gutierrez$^{12}$,
W.~G.~Hartley$^{1,7}$,
K.~Honscheid$^{32,33}$,
B.~Hoyle$^{8,9}$,
D.~J.~James$^{34}$,
M.~Jarvis$^{6}$,
K.~Kuehn$^{35}$,
M.~Lima$^{36,17}$,
H.~Lin$^{12}$,
M.~March$^{6}$,
P.~Melchior$^{37}$,
F.~Menanteau$^{19,20}$,
R.~Miquel$^{38,21}$,
A.~A.~Plazas$^{28}$,
K.~Reil$^{31}$,
A.~Roodman$^{24,31}$,
E.~Sanchez$^{25}$,
V.~Scarpine$^{12}$,
M.~Schubnell$^{30}$,
I.~Sevilla-Noarbe$^{25}$,
M.~Smith$^{39}$,
M.~Soares-Santos$^{12,40}$,
F.~Sobreira$^{41,17}$,
E.~Suchyta$^{42}$,
M.~E.~C.~Swanson$^{20}$,
G.~Tarle$^{30}$,
D.~Thomas$^{13}$,
A.~R.~Walker$^{11}$
\begin{center} (DES Collaboration) \end{center}
}
\vspace{0.2cm}
\parbox{\textwidth}{ \small
\textit{The authors' affiliations are shown in Appendix~\ref{sec:affiliations}. \\
}}
}}

\date{Accepted 10 May 2018}

\pubyear{2018}

\begin{document}
\label{firstpage}
\pagerange{\pageref{firstpage}--\pageref{lastpage}}
\maketitle

\begin{abstract}
Mapping the underlying density field, including non-visible dark matter, using weak gravitational lensing measurements is now a standard tool in cosmology. Due to its importance to the science results of current and upcoming surveys, the quality of the convergence reconstruction methods should be well understood. We compare three methods: Kaiser-Squires (KS), Wiener filter, and {\sc Glimpse}. KS is a direct inversion, not accounting for survey masks or noise. The Wiener filter is well-motivated for Gaussian density fields in a Bayesian framework. {\sc Glimpse} uses sparsity, aiming to reconstruct non-linearities in the density field. We compare these methods with several tests using public Dark Energy Survey (DES) Science Verification (SV) data and realistic DES simulations. The Wiener filter and {\sc Glimpse} offer substantial improvements over smoothed KS with a range of metrics. Both the Wiener filter and {\sc Glimpse} convergence reconstructions show a $12 $ per cent improvement in Pearson correlation with the underlying truth from simulations. To compare the mapping methods' abilities to find mass peaks, we measure the difference between peak counts from simulated $\Lambda$CDM shear catalogues and catalogues with no mass fluctuations (a standard data vector when inferring cosmology from peak statistics); the maximum signal-to-noise of these peak statistics is increased by a factor of 3.5 for the Wiener filter and 9 for {\sc Glimpse}. With simulations we measure the reconstruction of the harmonic phases; the phase residuals' concentration is improved $17$ per cent by {\sc Glimpse} and $18$ per cent by the Wiener filter. The correlation between reconstructions from data and foreground redMaPPer clusters is increased $18$ per cent by the Wiener filter and $32$ per cent by {\sc Glimpse}. 
\end{abstract}


%%%%%%%%%%%%%%%%%%%%%%%%%%%%%%%%%%%%%%%%%%%%%%%%%%

%%%%%%%%%%%%%%%%% BODY OF PAPER %%%%%%%%%%%%%%%%%%

\section{Introduction} \label{sec:intro}

Mass map reconstruction from weak gravitational lensing recovers the underlying matter distribution in the universe from measurements of galaxy shapes. Images of distant galaxies are deformed by the inhomogeneous matter distribution along the line of sight. Any matter can contribute to the lensing effect, making it a direct probe of non-visible dark matter.

Weak lensing, which takes advantage of the statistical power from many small distortions (that is, gravitational lensing induced ``shears''), is now a well established tool in constraining cosmology. The Dark Energy Survey (DES) has used the 2-point correlation function of shear to contribute to excellent constraints on cosmological parameters and models, including the nature of dark energy~\citep{des_y1_results}. Shear 2-point correlation functions have been used to constrain cosmology from many other survey datasets~(\citealt{kids_gamma_17},~\citealt{cfhtlens_kilbinger}). These methods use the shear measurements directly, as the shear can be related to the underlying matter distribution without needing to explicitly reconstruct mass maps.

A zero-mean Gaussian random field can be characterised entirely by its 2-point correlations. The matter density field in the early universe is expected to be highly Gaussian, a property which persists into the late universe for the large scales that were less affected by gravitational collapse. For the smaller scales at late times, non-linear gravitational collapse has led to a highly non-Gaussian density field. Much valuable information can be extracted from this non-Gaussianity, although this requires additional methods beyond 2-point statistics.

Popular proposed methods to extract this information include N-point statistics and higher order moments~\citep{cooray_bispectrum}, peak statistics (\citealt{peaks2010}, \citealt{des_peaks},~\citealt{peel_peaks},~\citealt{kids_peaks1},~\citealt{kids_peaks2}), and Minkowksi functionals (\citealt{kerscher_minkowski}, \citealt{petri_minkwoski_and_moments}). It is often either essential or convenient to apply these methods to the density field directly (rather than in the space of the shear measurements), thereby necessitating a reliable mass map reconstruction. 

Peak statistics are particularly promising, as peaks in the density field probe the non-Gaussian structure directly. Peaks can be identified from aperture mass maps, which are derived by convolving the shear data with a kernel, or from the reconstructed density field. The first approach has the advantage of having local noise, while the second is ``closer'' to the underlying density field and often has faster algorithms. Both methods often require simulations to provide a link between the theory and data, with the exception of proposed semi-analytic models (\citealt{lin_peaks_1}, \citealt{kids_peaks1}).

In addition to using mass maps for higher order statistics to constrain cosmological parameters and models, the mass maps can themselves be intrinsically useful. \cite{clerkin_lognnormal}, using the original DES Science Verification (SV) mass map, show evidence that the 1-point distribution of the density field is more consistent with Log-Normal than Gaussian. Combining mass maps with the spatial distributions of stellar mass, galaxies, or galaxy clusters allows the relationship between the visible baryonic matter and invisible dark matter to be studied. Using mass maps to constrain galaxy bias~\citep{changbias}, the relation between the distribution of galaxies and matter, can in turn aid cosmological probes other than weak lensing. Maps also enable simple tests for systematic error in the galaxy shape catalogues.

Since the first application of mass mapping methods to wide-field surveys with the Canada-France Hawaii Telescope Lensing Survey (CFHTLenS) data~\citep{cfhtlens_mass_map}, mass maps have been a standard product of large weak lensing surveys. In addition to DES, current surveys reconstructing the density field from weak lensing data include the Kilo-Degree Survey (Giblin et al. in prep.) and the Hyper Supreme-Cam Subaru Strategic Program (HSC-SSP)~\citep{hsc_map}. Mapping dark matter is key to the science goals of the future Euclid Mission~\citep{euclid_science} and the Large Synoptic Survey Telescope~\citep{lsst_science}.

DES is a ground based photometric galaxy survey, observing in the southern sky from the 4m Blanco telescope at the Cerro Tololo Inter-American Observatory (CTIO) in Chile with five photometric filters covering the optical and near-infrared spectrum using the Dark Energy Camera (\citealt{decam},~\citealt{more_than_de}). The SV data come from an initial run over a fraction of the final sky coverage, but to almost the full exposure time of the final survey. The sky coverage is still large, 139 deg$^2$, and the nearly full exposure~\citep{chang_sv_map} gives a galaxy density almost equal to what is expected after the complete 5 years of DES observations.

This paper uses the public DES SV data to compare the quality of mass mapping reconstruction methods. The maps are of the two-dimensional convergence, $\kappa$, a weighted projection of the density field in the foreground of the observed background galaxies. Recovering the convergence from the shear data is an ill-posed inverse problem, troubled by survey masks and galaxy ``shape noise''. 

This work follows on from that of~\cite{chang_sv_map} and~\cite{vikram_sv_map} in which the original DES SV mass map was created using the~\cite{KS93} method. In this paper we compare three quite different methods: Kaiser-Squires (KS); Wiener filtering~\citep{wiener1949extrapolation}; and {\sc Glimpse} (\citealt{leonard2014glimpse}, \citealt{lanusse_2016}), a sparsity-based reconstruction method. The Kaiser-Squires method is a direct inversion from shear to convergence, taking no account of missing data or the effect of noise. The Wiener filter and {\sc Glimpse} assume different prior knowledge about the underlying convergence to account for the effects of noise and missing data.

In Sec. 2 we describe the theoretical foundation for weak lensing mass mapping and the three different methods used for this work. In Sec. 3 we describe the DES SV shear data, the accompanying simulations, and the redMaPPer galaxy cluster catalogue. Foreground galaxy clusters are expected to trace the true density field, and therefore should be correlated with the convergence reconstruction. The different methods are also applied to realistic data simulations where the true convergence is known. In Sec. 4 we present our results on data and simulation, using various quality metrics for the reconstruction. On simulations these metrics are the Pearson correlation coefficient, the pixel root-mean-square error (RMSE), the variance of the 1-point distribution of pixel values, the phase residuals, and peak statistics. On data we compare the convergence reconstructions to the foreground galaxy clusters. We conclude in Sec. 5.

\section{Methodology}
\subsection{Weak gravitational lensing}

We can use measurements of the distortion of background galaxy shapes by weak gravitational lensing to learn about the mass distribution in the foreground without making many physical assumptions or relying on phenomenological models. For convenience, here we summarise some of the existing literature relevant for mass mapping from weak lensing (\citealt{bartelmann_schneider},~\citealt{kilbinger_cosmic_shear}). 

The weak lensing formalism follows photon paths along geodesics in a perturbed Friedmann-Robertson-Walker (FRW) metric. The perturbations are sourced by the density field of large scale structure. Throughout we assume that the perturbations are small, and that the measurements are made over a small enough patch of the sky that the sky geometry is Euclidian. Consistent with the Planck CMB results~\citep{ade2016planck} and motivated by inflationary theory, we assume that the global geometry of the universe is flat.

The density contrast, $\delta = (\rho - \bar{\rho})/\bar{\rho}$, of a pressureless fluid is related to the scalar gravitational potential perturbation, $\Phi$, through the Poisson equation,

\begin{equation}
\nabla^2 \Phi = \frac{3 H_0 \Omega_m}{2 a } \delta \ ,
\end{equation}

\noindent where $H_0$ is the present value of the Hubble parameter, $a$ is the cosmological scale factor, and $\rho$ and $\bar{\rho}$ are the local and mean density respectively.

For a flat universe, the lensing potential is given by

\begin{equation} \label{eq:lenspot}
\psi(\vv{\theta}, \omega) = 2 \int_0^\omega \mathrm{d} \omega' \Big[ \frac{\omega - \omega'}{\omega \omega'} \Big] \Phi(\vv{\theta} , \omega') \ ,
\end{equation}

\noindent where $\omega$ is the comoving distance.

The Born approximation assumes that the observed angle to a point, $\vv{\theta}$, deviates only a small amount from the true angle $\vv{\beta}$, so the change in distance of the photon's path is negligible. We can characterise the effect of lensing on the galaxies using the Jacobian of the transformation, $\mathcal{A}_{ij} = \partial \beta_i / \partial \theta_j$, which is decomposed into the functions $\kappa(\vv{\theta})$ and $\gamma(\vv{\theta}) = \gamma_1 + i \gamma_2$, and which is given by 

\begin{equation}
\begin{split}
\mathcal{A}  &= \Big( \delta_{ij} -  \frac{\partial^2 \psi (\vv{\theta})}{\partial \theta_i \partial \theta_j } \Big)
\\ &= \left( \begin{array}{cccc}
1 - \kappa - \gamma_1 & -\gamma_2  \\
-\gamma_2 &  1 - \kappa + \gamma_1  \end{array} \right) \ .
\end{split}
\end{equation}

\noindent Using the definition of the lensing potential and the Poisson equation, the convergence can be expressed as an integral over the density along the line of sight,

\begin{equation} \label{eq:Q}
\kappa(\vv{\theta}, \omega ) = \frac{3 H_0^2 \Omega_m}{2} \int_0^\omega \mathrm{d} \omega' \frac{\omega' (\omega - \omega')}{\omega} \frac{\delta(\vv{\theta}, \omega')}{a(\omega')} \ .
\end{equation}

\noindent For a distribution $n(\omega)$ of lensed galaxies, the lensing efficiency kernel is defined to be

\begin{equation} \label{eq:lensing_efficiency}
p(\omega') = \int^{\omega_\infty}_{\omega^\prime}  \left( \frac{\omega - \omega^\prime}{\omega}\right)n(\omega) \mathrm{d} \omega  \ ;
\end{equation}

\noindent this weights the contribution of the foreground density fluctuations to give the convergence weighted over the redshift distribution of source galaxies, 

\begin{equation} \label{eq:Q2}
\begin{split}
\kappa (\vv{\theta}) &= \int_0^\infty n(\omega) \kappa(\vv{\theta}, \omega ) \mathrm{d} \omega \\
 &= \frac{3 H_0^2 \Omega_m}{2} \int_0^\infty \mathrm{d} \omega' p(\omega')  \omega' \frac{\delta(\vv{\theta}, \omega')}{a(\omega')} \ .
\end{split}
\end{equation}

\noindent The shear, $\gamma( \vv{\theta} )$, which is assumed to be an observable in the weak lensing limit, is given by

\begin{equation} \label{eq:kappa}
\begin{split}
\gamma( \vv{\theta} ) = \frac{1}{\pi} \int_{\mathbb{R}^2} \mathrm{d}^2 \theta' \mathcal{D} ( \vv{\theta} - \vv{\theta}') \kappa(\vv{\theta}') \\
\mathrm{where} \; \mathcal{D} ( \vv{\theta} ) = - ( \theta_1 - i \theta_2 ) ^{-2} \ .
\end{split}
\end{equation}

\noindent For surveys where the integral is over large angles on the sky, this formulation breaks down, and requires a full treatment in spherical bases. \cite{wallis_projection} show that errors can be introduced at an $\mathcal{O}(1{\rm \ per\ cent})$ level for correlations between points at DES SV angular separation depending on the projection. All of the methods used here use the small angle approximation, and should suffer equally.

The real and imaginary parts of the shear $\gamma$ represent a chosen two dimensional coordinate system. In weak lensing, the observed ellipticity\footnote{Using the~\cite{bartelmann_schneider} equation 4.10 ellipticity definition for $\epsilon$.} of a galaxy $\epsilon_{obs}$ is related to the reduced shear $g$ plus the intrinsic ellipticity of the source galaxy $\epsilon_s$ through

\begin{equation} \label{eq:recuded_shear}
\begin{split}
\epsilon_{obs} &\approx g + \epsilon_s \\
& \ \  \mathrm{where} \ \ g = \frac{\gamma}{1 - \kappa} \ .
\end{split}
\end{equation}

\noindent The reduced shear is approximately the true shear, $g \approx \gamma $, in the weak lensing limit. This allows a standard definition of observed shear, $\gamma_{obs} = \epsilon_{obs}$, where the measurements are degraded by ``shape noise'', caused by the $\epsilon_s$ values of the observed galaxies:

\begin{equation} \label{eq:shape_noise}
\gamma_{obs} \approx \gamma +   \epsilon_s \ .
\end{equation}

\begin{figure}
\hspace*{0.15in}
\includegraphics[width=0.47\textwidth]{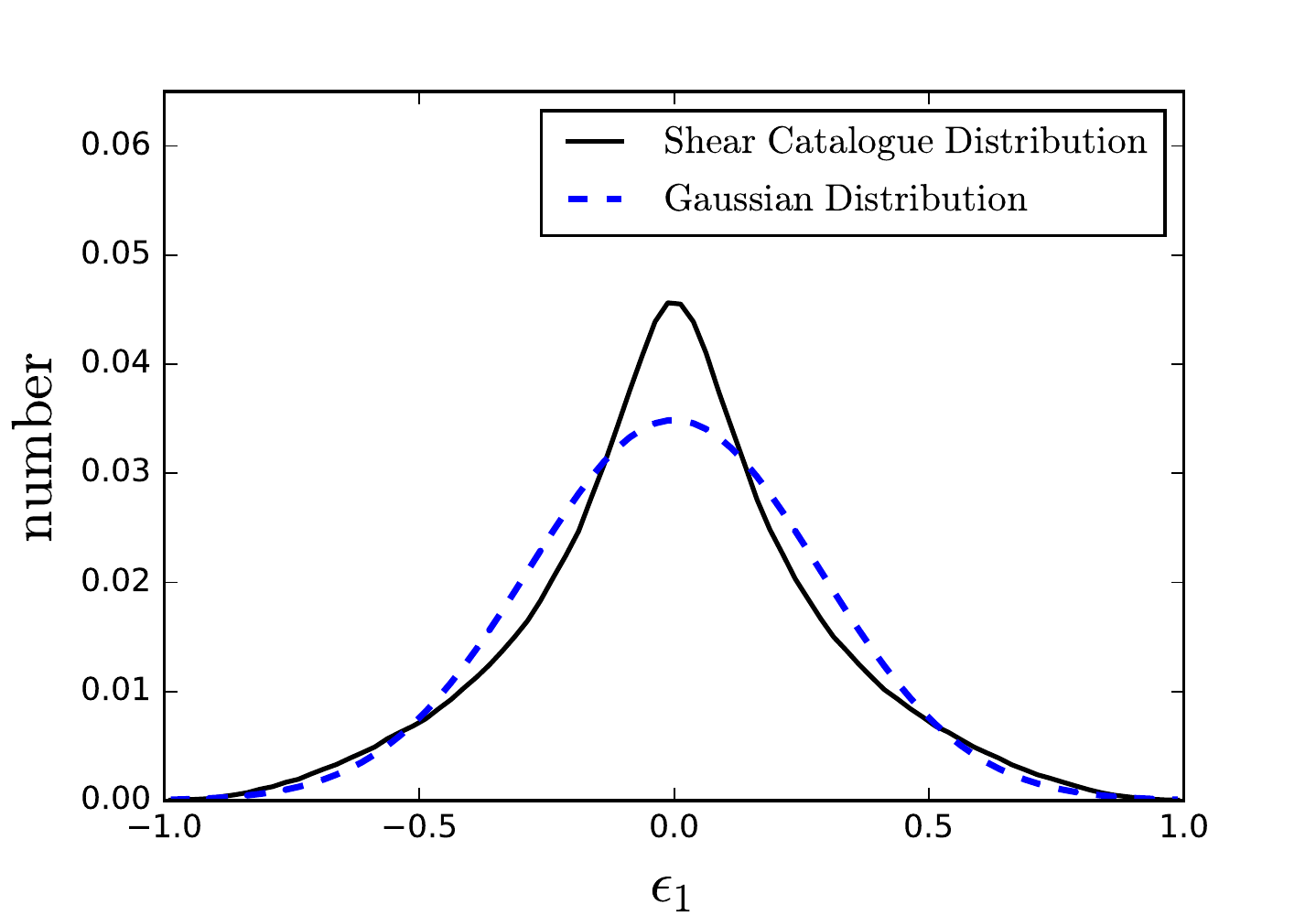}
\caption{\label{fig:epsilon} Distribution of the first component of ellipticity, $\epsilon_1$, from the selected SV catalogue. A Gaussian distribution with the same mean and standard deviation shows that the ellipticity distribution is not a true Gaussian, though the noise per pixel will be more closely Gaussian due to the central limit theorem.}
\end{figure}

\noindent The shape noise for a given galaxy is modelled as a randomly-drawn Gaussian variate, $\epsilon_s \sim G(0,\sigma_\epsilon)$, where $\sigma_\epsilon$ is estimated from data. The distribution of the ellipticity from the SV data in figure~\ref{fig:epsilon} is not an exact Gaussian, as the true distribution is the result of galaxy astrophysics, though a Gaussian still has properties that make it a good approximation. The Gaussian would be the maximum entropic, least informative, distribution for known mean and variance, and, by the central limit theorem, would be the correct distribution in the limit of large numbers of galaxies averaged in pixels.

It is possible to extend the simple Kaiser-Squires method (Sec.~\ref{sec:KS}) to use the reduced shear, $g$, for the mildly non-linear lensing regime when it is no longer appropriate to assume $g \approx \gamma $ (\citealt{schneider_seitz_1},~\citealt{schneider_seitz_2},~\citealt{seitz_scheider_finite_field}). This is also done by {\sc Glimpse} (Sec.~\ref{sec:glimpse}).

In matrix notation, the problem as given by equations~\ref{eq:kappa} and~\ref{eq:shape_noise} can be expressed as a linear model, with a data vector of observed shear measurements 

\begin{equation} \label{eq:matrix}
\mathbf{\gamma} = \mathbf{A} \mathbf{\kappa} + \mathbf{n} \ ,
\end{equation}

\noindent where $\mathbf{A}$ is a discretised version of equation~\ref{eq:kappa} and $\mathbf{n}$ is a noise vector due to shape noise (equation~\ref{eq:shape_noise}). The elements of the data vector can either correspond to the individual shear measurements or to measurements binned into angular pixels (in which case the noise vector would be the average noise in the pixel). 

The convergence need not be reconstructed with the same pixelisation as the shear measurements, giving $\kappa$ and $\gamma$ vectors of different length. Missing data due to survey masks would correspond to a shorter $\gamma$ vector; here one may wish to fill in the convergence in the masked region --- this is known as \textit{inpainting}. Different sized  $\kappa$ and $\gamma$ vectors result in a  non-square $\mathbf{A}$ matrix, potentially causing inversion problems.

\subsection{Kaiser-Squires reconstruction} \label{sec:KS}
\subsubsection{Theory}

The convergence-to-shear relationship, equation~\ref{eq:kappa}, is a convolution in the two dimensional angular plane. The two-dimensional Fourier transforms of the shear and convergence, defined for $\kappa$ as

\begin{equation}
\tilde{\kappa} ( \vv{k} ) = \int_{\mathbb{R}^2} \mathrm{d}^2 \theta \kappa(\vv{\theta}) \mathrm{exp} ( i \vv{\theta}  \cdot \vv{k} ) \ ,
\end{equation}

\noindent are related through an elementwise product via the convolution theorem

\begin{equation} \label{eq:kappa_fourier}
\tilde{\gamma} ( \vv{k} ) = \pi^{-1} \tilde{\mathcal{D}} ( \vv{k} ) \tilde{\kappa} ( \vv{k} ) \ ,
\end{equation}

\noindent where the Fourier transform of the kernel is given by

\begin{equation}
\tilde{\mathcal{D}} ( \vv{k} ) = \pi \frac{(k_1^2 - k_2^2 + 2 i k_1 k_2) }{\lvert \vv{k} \lvert^2} \ ;
\end{equation}

\noindent here $k_1$ and $k_2$ are the components of $ \vv{k}$. Using $\tilde{\mathcal{D}} \tilde{\mathcal{D}}^* = \pi^2$, equation~\ref{eq:kappa_fourier} can be rewritten:

\begin{equation}
\tilde{\kappa} ( \vv{k} ) = \pi^{-1} \tilde{\gamma} ( \vv{k} )  \tilde{\mathcal{D}}^* ( \vv{k} ) \; \; \; \; \mathrm{for} \; \; \vv{k} \neq \vv{0} \ .
\end{equation}

\noindent The inverse Fourier transform then returns the convergence reconstruction in configuration space~\citep{KS93}. 

The real and imaginary parts of the reconstruction are the E- and B-modes respectively, where $\kappa_{\rm{recon}} = \kappa_E + i\kappa_B$. In standard cosmology (equation~\ref{eq:kappa}), the convergence sourced by a real density field should be a pure E-mode. Errors, noise or other systematic effects can lead to B-mode contributions to the reconstruction. 

\subsubsection{Implementation}

In the matrix formulation of equation~\ref{eq:matrix}, this deconvolution corresponds to multiplying the Fourier space shear field with the inverse of $\mathbf{A}$ in Fourier space. For a case with no shape noise, that is

\begin{equation} \label{eq:noise_free}
\tilde{\mathbf{\gamma}} = \tilde{\mathbf{A}} \tilde{\mathbf{\kappa}} \ ,
\end{equation}

\noindent the Kaiser-Squires method is identical to using the inverse matrix

\begin{equation} \label{eq:ks_fourier_matrix}
\begin{split}
\big[ \tilde{\mathbf{A}}^{-1} \big]_{ij} &=  \frac{k_{1,i}^2 - k_{2,i}^2 - 2 i  k_{1,i} k_{2,i}}{{k_{1,i}^2 + k_{2,i}^2}} \delta_{ij} \\
&= \big[ \tilde{\mathbf{A}}^\dagger \big]_{ij}
\ ,
\end{split}
\end{equation}

\noindent where the Kronecker delta function, $\delta_{ij}$, relates the element-wise multiplication in Fourier space to a diagonal matrix operator, and $^\dagger$ is the conjugate transpose.

For the Kaiser-Squires inversion in configuration space, the $\mathbf{A}$ and $\mathbf{A}^{\dagger}$ matrices are not diagonal, and therefore are slower to compute. The discretisation of the underlying smooth shear field into finite configuration space makes the property $\mathbf{A} \mathbf{A}^{\dagger}  = \mathbf{I}$ inexact. As a result of these factors, we choose to implement the Kaiser-Squires reconstruction in Fourier space.

The shear due to lensing is much smaller than the shape noise, and not all places on the sky contain usable galaxies. Both the shape noise and the random sampling of background galaxies propagate error through this noisy reconstruction. Binning the shear measurements into larger pixels can reduce the shape noise per pixel and ensure that there are no empty pixels, but this comes at a loss of the small scale information and cannot deal with masks or the edges of the survey.

A smoothing filter is applied to the Kaiser-Squires reconstruction to reduce the noise. This will similarly lose any small scale structure, and especially suppress peaks in the convergence. In this work, matching~\cite{chang_sv_map}, we smooth the Kaiser-Squires maps with a Gaussian kernel. The standard deviation scale, $\sigma_{smooth}$, of this Gaussian kernel is free to be chosen, where $\sigma_{smooth} = 0$ corresponds to standard, unsmoothed Kaiser-Squires.

\subsection{Wiener Filter}
\subsubsection{Theory} \label{sec:wiener_theory}

The Wiener filter is the linear minimum-variance solution to linear problems of the type in equation~\ref{eq:matrix}, where the noise is uncorrelated. The Wiener filter reconstruction (\citealt{lahav1993wiener},~\citealt{zaroubi_wiener}) is given by 

\begin{equation} \label{eq:wiener}
\begin{split}
\mathbf{\kappa}_W &= \mathbf{W} \mathbf{\gamma} \\
\mathbf{W} &= \mathbf{S}_\kappa \mathbf{A}^\dagger \big[ \mathbf{A} \mathbf{S}_\kappa \mathbf{A}^\dagger + \mathbf{N} \big]^{-1} \ .
\end{split}
\end{equation}

\noindent Here $\mathbf{S}_\kappa$ and $\mathbf{N}$ are the signal and noise covariance matrices respectively, which are $\langle \mathbf{\kappa} \mathbf{\kappa}^\dagger \rangle $ and $\langle \mathbf{n} \mathbf{n}^\dagger \rangle$ for this problem. 

This filter is the linear minimum-variance solution, as  $\mathbf{W}$ is a linear operator that minimises the variance

\begin{equation} \label{eq:wiener_rmse}
\langle (\mathbf{W \gamma} - \mathbf{\kappa})^\dagger (\mathbf{W \gamma} - \mathbf{\kappa}) \rangle \ .
\end{equation}

\noindent If the chosen prior on $\mathbf{\kappa}$ does not constrain the reconstruction, so that $\mathbf{S}_\kappa^{-1} \to 0$~\citep{simon_2009}, or if the data are noise-free, $\mathbf{N} = 0$, then the linear minimum variance filter becomes the Kaiser-Squires reconstruction. Setting $\mathbf{S}_\kappa^{-1} \to 0$ is equivalent to removing the signal prior in the following Bayesian framework.

From a different starting point, for the Wiener posterior we begin by assuming a Gaussian likelihood~\citep{jasche2015matrix}

\begin{equation}
Pr( \mathbf{\gamma} | \mathbf{\kappa} ) = \frac{1}{\sqrt[]{(\mathrm{det} 2 \pi \mathbf{N})}} \mathrm{exp} \Big[ - \frac{1}{2} ( \mathbf{\gamma} - \mathbf{A} \mathbf{\kappa} )^\dagger \ \mathbf{N}^{-1} ( \mathbf{\gamma} - \mathbf{A} \mathbf{\kappa} )  \Big] \ , 
\end{equation}

\noindent where it is assumed that $\mathbf{N}$ is known and the noise is both uncorrelated and Gaussian, as assumed in equation~\ref{eq:matrix}. Intrinsic alignments of clustered galaxies will violate this uncorrelation condition.

The prior on the convergence is that of a Gaussian random field, which is applicable for the density field on large scales at late times, 

\begin{equation}
Pr( \mathbf{\kappa} | \mathbf{S}_\kappa ) = \frac{1}{\sqrt[]{(\mathrm{det} 2 \pi \mathbf{S}_\kappa)}} \mathrm{exp} \Big[ - \frac{1}{2} \mathbf{\kappa}^\dagger \ \mathbf{S}_\kappa^{-1} \mathbf{\kappa}  \Big] \ .
\end{equation}

Using Bayes' theorem and the fact that $Pr(\mathbf{\gamma} | \mathbf{S}_\kappa, \mathbf{\kappa} ) = Pr(\mathbf{\gamma} | \mathbf{\kappa} )$, the full posterior is given by

\begin{equation} \label{eq:wienerpost}
\begin{split}
Pr( \mathbf{\kappa} | \mathbf{S}_\kappa, \mathbf{\gamma} ) &= \frac{  Pr( \mathbf{\gamma} | \mathbf{\kappa} ) Pr( \mathbf{\kappa} | \mathbf{S}_\kappa ) }{Pr( \mathbf{\gamma} )} \\
 &\propto \frac{1}{\sqrt[]{(\mathrm{det} 2 \pi \mathbf{S}_\kappa)}} \frac{1}{\sqrt[]{(\mathrm{det} 2 \pi \mathbf{N})}} \times \\ &\ \ \ \  \mathrm{exp} \Big[ - \frac{1}{2} \mathbf{\kappa}^\dagger \mathbf{S}_\kappa^{-1} \mathbf{\kappa}  - \frac{1}{2} ( \mathbf{\gamma} - \mathbf{A} \mathbf{\kappa} )^\dagger \mathbf{N}^{-1} ( \mathbf{\gamma} - \mathbf{A} \mathbf{\kappa} ) \Big] \\
 &\propto \mathrm{exp} \Big[  - \frac{1}{2} ( \mathbf{\kappa} - \mathbf{W} \mathbf{\gamma} )^\dagger (  \mathbf{S}_\kappa^{-1} + \mathbf{A} \mathbf{N}^{-1} \mathbf{A}^\dagger ) (\mathbf{\kappa} - \mathbf{W} \mathbf{\gamma}) \Big] \ ,
\end{split}
\end{equation}

\noindent where $\mathbf{W}$ is the Wiener filter, so the \textit{maximum a posteriori} (MAP) solution is that of the Wiener reconstruction.

The choice of Gaussian prior is physically motivated for the large, linear scales of the density field (see section~\ref{sec:intro}); alternative prior distributions can be used to give different $\mathbf{\kappa}$ posterior distributions that can be maximised or from which samples can be drawn~\citep{schneider_probabilistic_map}. Recent work by~\cite{bohm_lognormal} proposes the use of a Log-Normal prior distribution. This appears to fit the $\mathbf{\kappa}$ distribution from simulations (figure~\ref{fig:kappa_hist}) and data~\citep{clerkin_lognnormal} better than Gaussian, but, unlike the Wiener filter, lacks an analytic MAP solution.  

If the aim of the reconstruction is to infer cosmology from the non-Gaussian component of the density field, the Wiener filter may not be the ideal method for mass map recovery. The small scale modes with less power are often suppressed, losing the peak structure. Qualitatively it can be thought of as either the Gaussian prior being inappropriate or as the linear filter being insufficient.

\subsubsection{Implementation}

Using the exact Fourier space property $\tilde{\mathbf{A}}^{-1} = \tilde{\mathbf{A}}^{\dagger}$ we rewrite equation~\ref{eq:wiener} as

\begin{equation}
\begin{split}
\tilde{\mathbf{\kappa}}_W &= \tilde{\mathbf{A}}^{-1} \tilde{\mathbf{S}}_\gamma \big[ \tilde{\mathbf{S}}_\gamma + \tilde{\mathbf{N}} \big]^{-1} \tilde{\mathbf{\gamma}} \\ &= \tilde{\mathbf{A}}^{-1} \tilde{\mathbf{\gamma}}_{W} \\
&=\tilde{\mathbf{A}}^{\dagger} \tilde{\mathbf{\gamma}}_{W} \ , 
\end{split}
\end{equation}

\noindent where we have used $\tilde{\mathbf{A}} \tilde{\mathbf{S}}_\kappa \tilde{\mathbf{A}}^\dagger =  \langle \tilde{\mathbf{A}} \tilde{\mathbf{\kappa}} \tilde{\mathbf{\kappa}}^\dagger \tilde{\mathbf{A}}^\dagger \rangle  = \langle \tilde{\mathbf{\gamma}} \tilde{\mathbf{\gamma}}^\dagger \rangle = \tilde{\mathbf{S}}_\gamma$. This shows that applying the Wiener filter to the shear to recover $\mathbf{\gamma}_W$ and then applying the Kaiser-Squires inversion in Fourier space is equivalent to directly calculating the Wiener filter of the convergence.

In configuration space, the noise covariance matrix is given by

\begin{equation} \label{eq:noise_cov}
\big[ \mathbf{N} \big]_{ij} = \frac{2 \sigma^2_\epsilon}{p_i} \delta_{ij} \ ,
\end{equation}

\noindent where $p_{i}$ is the galaxy count per pixel. Empty pixels in the masked region have infinite variance, absorbing the mask into a special case of the Wiener filter denoising.

The signal properties for a Gaussian random field are constrained entirely by the mean and the signal covariance matrix, which in harmonic space is identical to the power spectrum. The cosmological principle implies that the angular distribution of a field on the sky is statistically isotropic, so the angular power spectrum, $C_\ell$, can contain all the 2-point statistical information. The angular power spectrum of the physical shear E-mode shear signal is defined as

\begin{equation} \label{eq:cl}
\begin{split}
C_{\ell, E} &= \frac{1}{2 \ell + 1} \sum_{m = -\ell}^{+\ell} \langle |a_{\ell m, E}|^2\rangle \\
&= C_{\ell, \kappa}\ ,
\end{split}
\end{equation}

\noindent where $a_{\ell m}$ are the spherical harmonic coefficients and the brackets $\langle \  \rangle$ average over realisations of the signal. The second equality assumes the flat sky approximation for high\footnote{We omit a prefactor which goes as $1 - \mathcal{O}(\ell^{-2})$ for high $\ell$.}$~\ell$.

We generate a theoretical power spectrum using the Limber approximation with the {\sc Cosmosis} package~\citep{cosmosis} with our prior fiducial cosmological parameters: $\Omega_m = 0.286,\ \Omega_\Lambda = 0.714,\ \Omega_b = 0.047,\ h = 0.7,\ \sigma_8 = 0.82,\ n_s = 0.962 \rm{\ and\ } w = -1$. We use a background galaxy distribution defined from equation~\ref{eq:nz}, and shown in figure~\ref{fig:pz}.

It is commonly asked whether it is reasonable to assume cosmological parameters in the map reconstruction, if the maps are then used to infer cosmological parameters. Though we assume a specific set of cosmological parameters, it would still be possible to use the maps for cosmological parameter estimation, from peak statistics for example, if the same prior is used on the simulations and the data identically. If simulations are not used, the power spectrum can be jointly inferred from the data~\citep{jasche2015matrix} using Gibbs sampling.

In order to generate the power spectrum in flat Fourier space, rather than on the curved sky, we again use a flat sky approximation

\begin{equation}
k^2_{\theta} P(k_{\theta}) =  \Bigg( \frac{\mathcal{N}}{2 \pi} \Bigg)^2 \ell ( \ell + 1) C_\ell  \ ,
\end{equation}

\noindent adapted from~\cite{extended_limber}, where $\mathcal{N}$ is the total number of pixels in the map, $k_{ \theta}$ is the magnitude of the projected Fourier mode, and where we have defined our projected angular power spectrum as

\begin{equation} \label{eq:power_shear}
P(k_{\theta}) \delta( \mathbf{k}_{\theta} - \mathbf{k}^\prime_{\theta})= \langle \ \tilde{\mathbf{\gamma}} (\mathbf{k}_{ \theta} ) \ \tilde{\mathbf{\gamma}}^\dagger  (\mathbf{k}_{ \theta}) \ \rangle \ .
\end{equation}

\noindent The largest scale mode is $\ell = 20.51$, which corresponds to an angular separation of $17.55 \rm{\ deg}$.

Though the signal covariance matrix is diagonal in harmonic space (equation~\ref{eq:power_shear}), and the independent noise has covariance which is diagonal in configuration space (equation~\ref{eq:noise_cov}), there is no natural basis in which both are sparse. Inversion of dense matrices to evaluate the Wiener filter is bypassed using the algorithm presented in~\cite{elsner_wiener}, where an additional messenger field is used to pass information between harmonic and configuration space, iteratively converging to the Wiener filter solution. 

These messenger field methods were extended by~\cite{jasche2015matrix} to draw Markov chain Monte Carlo (MCMC) samples from the whole Wiener posterior (equation~\ref{eq:wienerpost}). The first application of messenger field methods to weak lensing data was by~\citeauthor{alsing_shear_maps}~(\citeyear{alsing_power_spectrum},~\citeyear{alsing_shear_maps}), who drew samples from the Wiener posterior and generated Wiener filtered shear (not convergence) maps from CFHTLenS data. By comparison, in this work we do not sample from the Wiener posterior; instead, we use the original messenger field algorithm of~\cite{elsner_wiener} to calculate the Wiener filter reconstruction of the convergence map from DES SV shear data and simulations. 

\subsection{Sparsity reconstruction}~\label{sec:glimpse}
\subsubsection{Theory}
Consider the coefficients $\mathbf{\alpha}$ of the decomposition of a signal $\mathbf{x}$ in a representation space (or ``dictionary'') $\mathbf{\Phi}$, so that $\mathbf{x} = \mathbf{\Phi} \mathbf{\alpha}$. Example dictionaries include the Fourier transform or wavelet transforms. Assuming a sparse prior on the signal $\mathbf{x}$ in the dictionary $\mathbf{\Phi}$ means that its representation $\mathbf{\alpha}$ is expected to be sparse, that is, with most of the coefficients equal to 0~\citep{Starck_sparsity}. A simple example is a cosine function signal and a Fourier transformation; in this sparse basis only two coefficients have a non-zero value (corresponding to the frequency of the cosine function).

Formally most signals cannot strictly be made sparse, and are merely compressible with a choice of an appropriate transformation, such as a wavelet transform (\citealt{Starck_sparsity},~\citealt{leonard2014glimpse}). For a compressible signal the magnitude-ordered sparse coefficients, $\alpha_i$, are expected to have exponential decay and therefore to have a Laplace distribution~\citep{Tibshirani94regressionshrinkage}. 

Consider a generic linear inverse problem of the form $\mathbf{y} = \mathbf{Ax} + \mathbf{n}$. A robust estimate of the signal $\mathbf{x}$ can be recovered by solving the (``LASSO'') optimisation problem

\begin{equation}
\underset{\mathbf{\alpha}} \argmin \ \  || \mathbf{y} - \mathbf{A} \mathbf{\Phi} \mathbf{\alpha} ||^2_2 + \lambda || \mathbf{\alpha} ||_1  \ ,
\end{equation}

\noindent where $\lambda$ is a Lagrangian multiplier~\citep{Tibshirani94regressionshrinkage}. Here the first term corresponds to a $\chi^2$ minimisation, ensuring fidelity of the signal reconstruction, while the second is the sparsity-promoting regularisation term.

We can include non-constant noise variance by weighting the first $\chi^2$ according to the variance. If the noise variance is included in the $\chi^2$ term, the $\lambda$ value can be interpreted as a signal-to-noise level in the transformed (e.g. wavelet) space.

The second term does not use the Euclidan $l_2$ norm, but instead uses the sparsity-promoting $l_1$ norm, defined as

\begin{equation}
|| \mathbf{\alpha} ||_1 = \sum_i | \alpha_i | \ .
\end{equation}

\noindent These methods are non-linear, so it can be difficult to derive properties analytically. With realistic simulations of the data and true signal, the value for $\lambda$ can be chosen to maximise some success metric. This is analogous to selecting a theoretical power spectrum for the Wiener filter, or a smoothing scale for Kaiser-Squires.

Sparse recovery methods are non-linear and are not necessarily formulated in the Bayesian framework of the Wiener filter. The Wiener filter reconstruction is that which maximises the Wiener posterior, which is known analytically provided the noise and signal are Gaussian with known covariance. However, one may make a frequentist estimate of the error of the sparse reconstruction by propagating the noise properties of the data using bootstrapping or Monte Carlo techniques.

\subsubsection{Implementation/\sc{Glimpse}}

The choice of dictionary depends on the structures contained in the signal. Theory of structure formation in the universe predicts the formation of quasi-spherical halos of bound matter. It is standard practice to represent the spatial distribution of matter in halos with spherically symmetric Navarro-Frenk-White~\citep{nfw} or Singular Isothermal Sphere profiles. Coefficients of Isotropic Undecimated Wavelets~\citep{Starck_sparsity} in two dimensions are well suited to the observed convergence of a dark matter halo. The wavelet transform used in the {\sc Glimpse} algorithm is the starlet~\citep{starck2007undecimated}, which can represent positive, isotropic objects. 

The sparsity prior in the starlet basis enforces a physical model that the matter field is a superposition of spherically symmetric dark matter halos. This is not wholly correct, but is an approximation which is true for the non-linear regime in the standard model of structure formation, similarly to how the assumption of Gaussianity holds in the linear regime. On large scales, where the density field is expected to be Gaussian, the {\sc Glimpse} sparsity prior is less appropriate.

The {\sc Glimpse} algorithm aims to solve the optimisation problem

\begin{equation} \label{eq:optimisation}
\begin{split}
\hat{\mathbf{\kappa}} = \underset{\mathbf{\kappa}} \argmin \ \  ||\mathbf{N}^{-\frac{1}{2}} \big[ \mathbf{\gamma} &- \mathbf{T}^{\dagger} \hat{\mathbf{A}} \mathbf{F} \mathbf{\kappa} \big] ||^2_2 \\ &+ \lambda || \mathbf{\omega} \mathbf{\Phi}^\dagger \mathbf{\kappa} ||_1  +   i_{\rm{Im}(\mathbf{\kappa}) = 0} \ ,
\end{split}
\end{equation}

\noindent where $\mathbf{F}$ is the Fourier transform matrix, $\mathbf{T}$ is the Non-equispaced Discrete Fourier Transform (NDFT) matrix, $\hat{\mathbf{A}}$ is defined in equation~\ref{eq:ks_fourier_matrix}, $\mathbf{\omega}$ is a diagonal matrix of weights, and $\mathbf{\Phi}^\dagger$ is the inverse wavelet transform. The indicator function $i_{\rm{Im}(\cdot) = 0}$ (defined in Appendix~\ref{sec:indicator}) in the final term imposes realness on the reconstruction (no B-modes). The use of NDFT allows the first term to perform a forward fitted Kaiser-Squires-like step without binning the shear data, allowing the smaller-scales to be retained in the reconstruction. The full algorithm, including the calculation of the weights, is described in Sec. 3.2 in \cite{lanusse_2016}.

Though the problem presented in equation~\ref{eq:optimisation} is an optimisation using the shear data $\gamma$, in fact it is the reduced shear (equation~\ref{eq:recuded_shear}) that {\sc Glimpse} uses to recover $\kappa$~\citep{lanusse_2016}. As an extension, the {\sc Glimpse} algorithm can also perform the joint reconstruction with reduced shear and flexion, a third-order weak gravitational lensing effect~\citep{bacon_flexion} (although no flexion data are available for our galaxy shear catalogue). 

As the prior knowledge in this reconstruction relates to the quasi-spherical clustering of bound matter, enforced through a sparsity prior in starlet space, this method should better reconstruct the smaller scale non-Gaussian structure than the Wiener filter. 

\section{Data and Simulations}

\subsection{Dark Energy Survey Science Verification Data}

The shear data are from the 139 deg$^2$ SPT-E field of the public DES SV data. This initial test data set was taken during an observing run before the official start of the full science survey. The galaxy catalogue comes from the SVA1 (Science Verification) Data Release\footnote{http://des.ncsa.illinois.edu}. Due to changes to the catalogues before final release (more galaxy shear measurements are now available to us), the catalogue used in this work is not identical to that used by~\cite{chang_sv_map}, even when the same data selections are made. All maps are therefore new, and slightly different to the previously published SV map.

\begin{figure}
\includegraphics[width=0.47\textwidth]{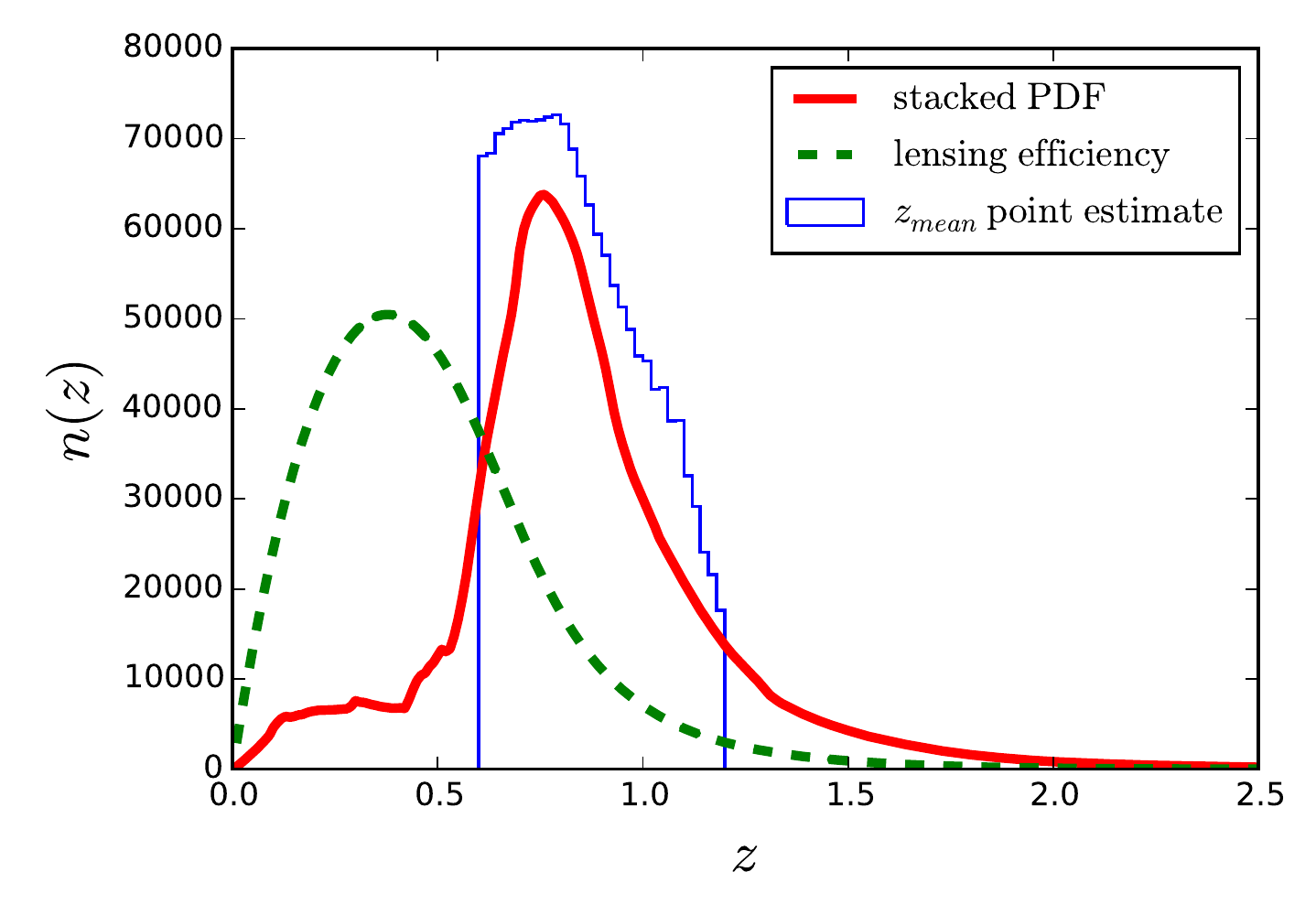}
\caption{\label{fig:pz} The redshift distribution from BPZ of the selected background galaxies with $0.6<z_{mean} <1.2$. The blue solid histogram is of the galaxies' point estimate mean redshifts in bins of $\Delta z = 0.02$. The red line is the stacked redshift probability density function (PDF) of all selected galaxies. The green dashed line is the lensing efficiency (equation~\ref{eq:lensing_efficiency}) of the background galaxies.}
\end{figure}

The photometric redshifts from five optical filters (\textit{grizY}) were estimated using the Bayesian Photometric Redshifts (BPZ) code (\citealt{bpz1},~\citealt{bpz2}, \&~\citealt{bonnett_photoz}). The final median depth estimates are $g\sim$24.0, $r\sim$23.0, $i\sim$23.0 and $z\sim$22.4 (10-$\sigma$ galaxy limiting magnitude). The ``background galaxies'', the ones from which the shear is measured, are taken in the range $0.6 < z_{mean} < 1.2$. The $z_{mean}$ value for each galaxy is the mean of the posterior probability distribution function (PDF) estimated using the BPZ code. The PDF for each galaxy is very broad, giving a total stacked PDF of background galaxies that extends beyond the $[0.6,1.2]$ redshift range, as can be seen in figure~\ref{fig:pz}.

Using the $\rm \tt ngmix$ shape catalogue, we apply a selection of ${\rm \tt sva1\_flag}= 0$ \& ${\rm \tt ngmix\_flag} = 0$ to obtain galaxies with a well-measured shear. The  $\rm \tt ngmix$ catalogue contains corrections to measurement bias, in the form of ``sensitivities'', which can be applied to a weighted ensemble of hundreds or thousands of galaxies, but which cannot be applied per galaxy (which is not ideal for mass mapping). The structure of equation~\ref{eq:kappa} implies that a multiplicative shear bias would lead to a convergence amplitude bias. Under the assumption that multiplicative shear bias will not vary across the survey area, we correct all measured ellipticities by the same debiasing factor

\begin{equation}
\epsilon_{obs, i} =\epsilon_{measured, i} \times \bar{s}^{-1} \ ,
\end{equation}

\noindent where $i$ is a galaxy index and $\bar{s} \ ( \approx 0.82 )$ is the mean sensitivity correction from all galaxies in our $\rm \tt ngmix$-selected catalogue. The total number of galaxies after the redshift and shape measurement selection is $1,628,663$.

For the Kaiser-Squires reconstruction, the shear measurements are binned into angular pixels in a $256 \times 256$ map, with average pixel size of $4.11 \mathrm{\ arcmin}$, using a sinusoidal projection with a centre at RA=$71.0$ deg. This is similar to the 5 arcmin pixel scale of the original~\cite{chang_sv_map} map. The choice of central RA for Kaiser-Squires is to minimise the mask in the square projection, which is a large source of systematic error. For the Wiener filter, where the mask is taken into account, the shear measurements are also binned into angular pixels in a $256 \times 256$ map, but sinusoidally projected with a central RA=$81.3$ deg, to make the square maximally isotropic. The {\sc Glimpse} algorithm does not bin the input shear measurements, but requires a pixel scale for the reconstruction, which we set as $3 \rm{\ arcmin}$ using its gnomonic projection centred on RA=$76.95$ deg and DEC=$-52.23$ deg.

% \begin{figure}
% \includegraphics[width=0.47\textwidth]{cropped_number_epsilon.png}
% \caption{\label{fig:epsilon} Distribution of the first component of ellipticity, $\epsilon_1$, from the selected SV catalogue. The second component, $\epsilon_2$, is not plotted, but shows the same distribution.}
% \end{figure} 

\subsection{redMaPPer Clusters} \label{sec:redmapper}

Groups and clusters of galaxies are expected to trace the highest density regions in the foreground. They are luminous objects that correspond to regions of highly non-linear growth, where the density field has deviated from Gaussianity.

The public redMaPPer cluster catalogue~\citep{redmapper} used the redMaPPer algorithm to optically identify clusters and to estimate each cluster's richness, $\lambda_{RM}$. The richness is defined as the sum of the membership probabilities over all galaxies within a scale radius (chosen to minimise the scatter in the mass-richness relation); it gives an estimate for the number of galaxies in a cluster. Cluster mass is expected to scale approximately linearly with richness. The redshift uncertainty is excellent, around $\sigma_{z} / (1+z) \sim 0.01$, due to the clusters containing large numbers of well modelled, red galaxies. The public redMaPPer catalogue used in this work contains only clusters with $\lambda_{RM} \geqslant 20$, so that the clusters with less certainty of detection and characterisation are not used.

\begin{figure*}
\centering
\hspace*{-0.9in}
\includegraphics[width=1.25\textwidth]{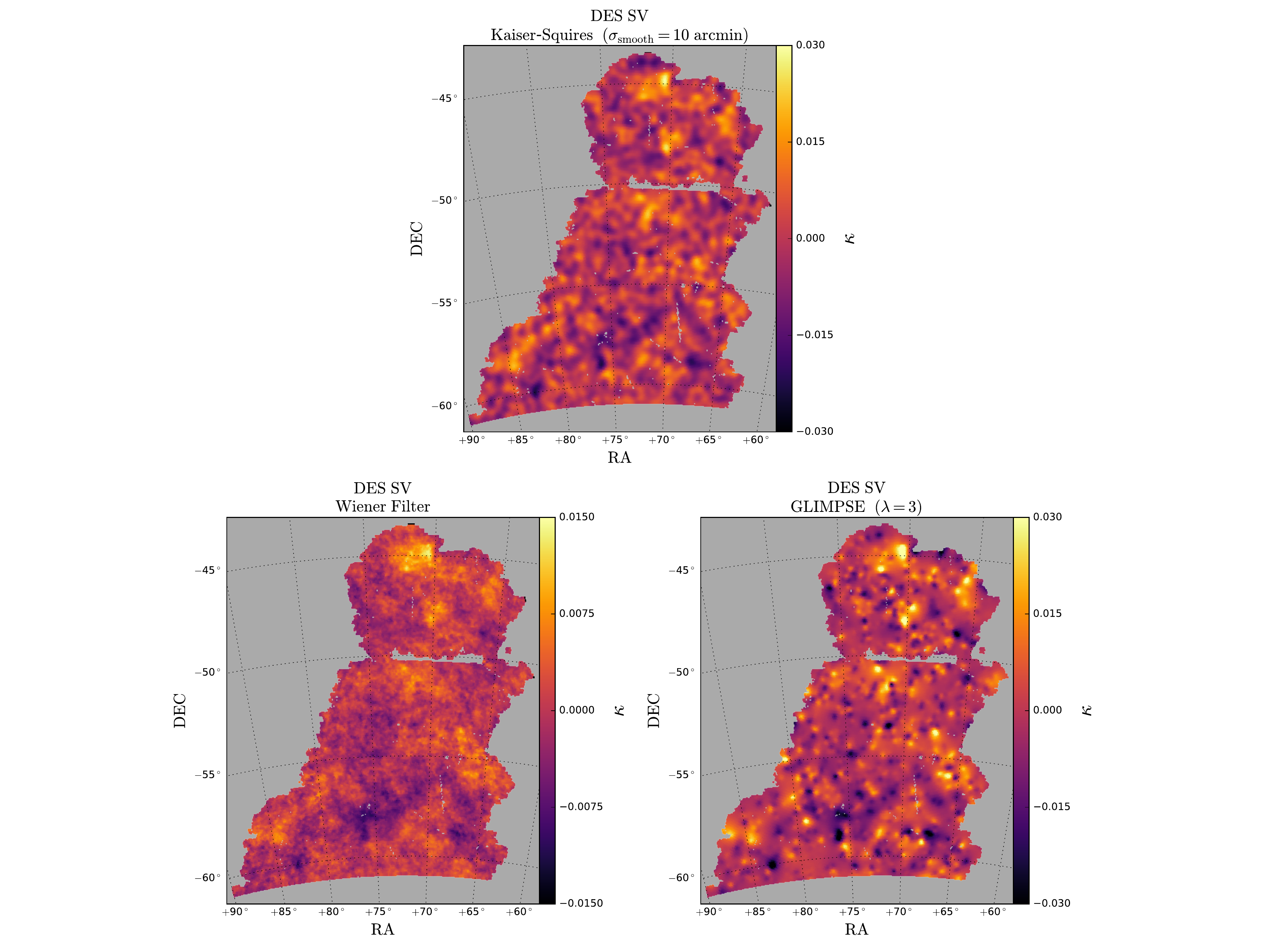}
\caption{\label{fig:desmap} The convergence ($\kappa$) map reconstructions using the DES SV shear data with the three different methods. \textit{Top panel}: Kaiser-Squires reconstruction with a smoothing scale $\sigma_{smooth} = 10 \rm{\ arcmin}$. \textit{Right panel}: The {\sc Glimpse} reconstruction with a regularisation parameter $\lambda = 3.0$. Both tuning parameters were chosen to maximise the Pearson correlation coefficient $r$ when tested on simulations (See Sec.~\ref{sec:pearson}). \textit{Left panel}: The Wiener filter reconstruction. \textbf{Note} that the colour scale for the Wiener filter is less than that for the other reconstructions, as the pixel values are closer to zero.}
\end{figure*} 

\begin{figure*}
\centering
\includegraphics[width=.93\textwidth]{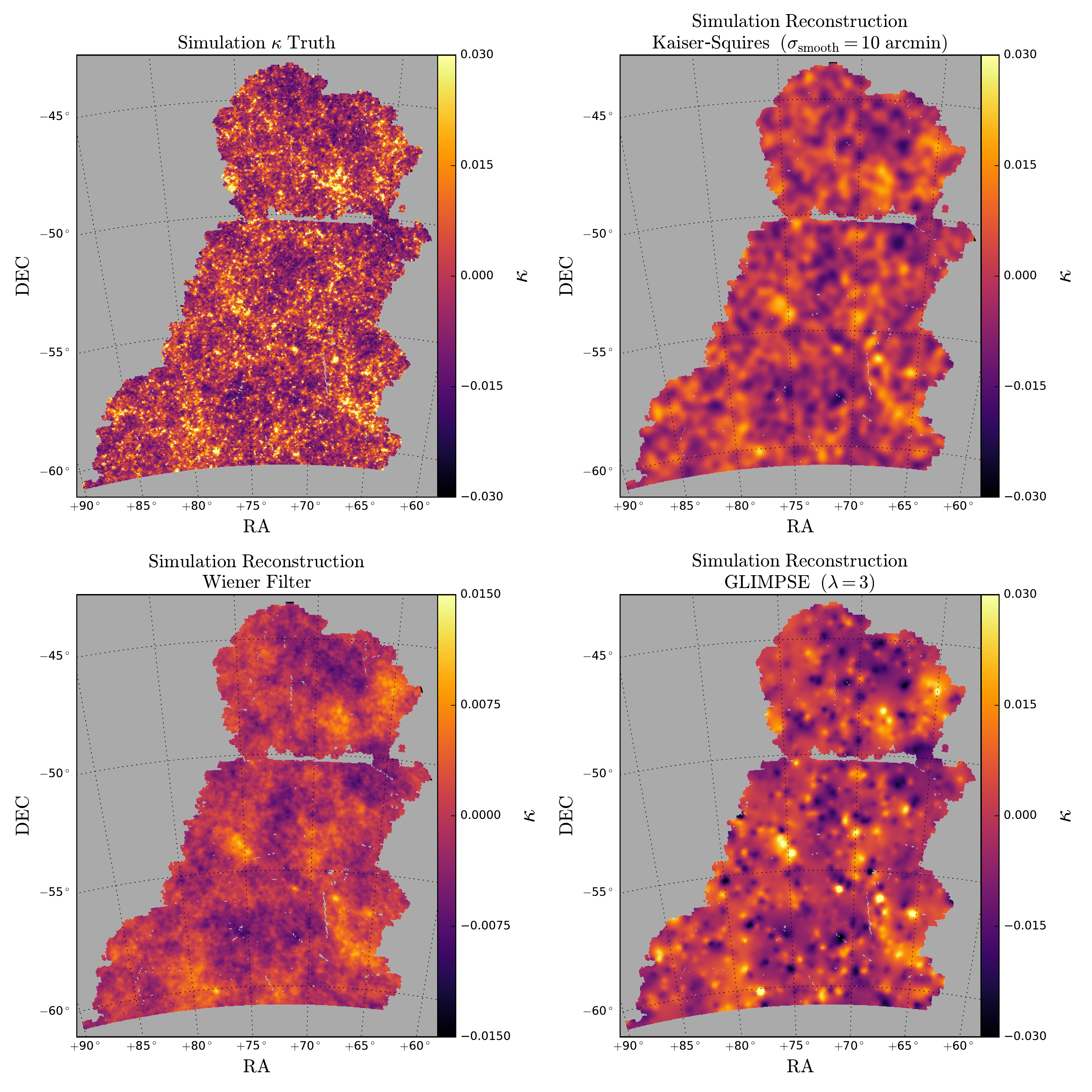}
\caption{\label{fig:simmap} The \textit{top left} panel is an example of a true convergence map, $\kappa^{\rm{truth}}$, from simulation. The \textit{top right} panel is the Kaiser-Squires reconstruction with a smoothing scale $\sigma_{smooth} = 10 \rm{\ arcmin}$. The \textit{bottom right} panel is the {\sc Glimpse} reconstruction with regularisation parameter $\lambda = 3.0$.  Both tuning parameters were chosen to maximise the Pearson correlation coefficient $r$ when tested on simulations (see Sec.~\ref{sec:pearson}). The \textit{bottom left} panel is the Wiener filter reconstruction. \textbf{Note} that the colour scale for the Wiener filter is less than that for the other reconstructions, as the pixel values are closer to zero.}
\end{figure*} 

\subsection{Simulations}

To compare the reconstructions between different methods, we use a simulated catalogue with a known true convergence. We use a set of N-body simulations developed for the DES collaboration and designed to be representative of the DES data~\citep{busha_bcc}. The simulations used are N-body light cones composed from three boxes ($1400^3$, $2048^3$, and $2048^3$ particles in boxes of comoving length 1050 Mpc/h, 2600 Mpc/h, and 4000 Mpc/h respectively). The cosmological parameters for the simulations are: $\Omega_m = 0.286$, $\Omega_\Lambda = 0.714$, $\Omega_b = 0.047$, $\sigma_8 = 0.82$, $h_0 = 0.7$, $n_s = 0.96$, $w = -1$. We apply a mask to match the SV data.

Source galaxies have randomly-assigned positions in the simulations, as correlation between the background galaxy positions and the weak lensing shear signal is expected to be negligible. The simulated catalogues contain the lensing matrix components, $\mathcal{A}_{ij}$, for each galaxy, calculated with the ray-tracing code $\rm \tt CALCLENS$~\citep{becker_calclens}. This provides the true $\kappa$ and $\gamma$ per galaxy, from which we derive the reduced shear. The shape noise due to the intrinsic ellipticities of the source galaxies, $\epsilon_s$, is simulated by adding an ellipticity component to the reduced shear. Each noise realisation is generated from the data by randomly exchanging the ellipticity values between galaxies in the catalogue to remove the weak lensing signal and leave the shape noise.

We attempt to match the redshift distribution of the simulated galaxies to the observed redshift distribution, $n(z)$. We use the stacked posterior probability density functions of individual galaxy redshifts from the selected data catalogue (figure~\ref{fig:pz}), giving an estimate of the true underlying distribution. This assumes that

\begin{equation} \label{eq:nz}
n(z) = \sum_i p_i(z) \ ,
\end{equation}

\noindent where $p_i(z)$ are the individual probability distributions for the galaxies from BPZ. This is not necessarily exact, due to errors in $p_i (z)$ per galaxy \citep{leistedt_nz}, but is a reasonable choice for a simulated catalogue. Using rejection sampling in bins of $\Delta z = 0.02$ we select galaxies with a probability equal to the ratio between the desired $n(z)$ from the data and the distribution in the simulation. One typical simulated catalogue contained $1,629,024$ galaxies, slightly different to the data catalogue due to the sampling scheme, but with the desired $n(z)$.

\section{Results}

To ensure that the mass map tests are consistent with different output formats, all maps were converted onto a spherical pixelisation using ${\rm \tt HEALPix}$~\citep{healpix}. A ${\rm \tt HEALPix}$ map comprises twelve subdivisions on the sphere, which are then each partitioned into ${\rm \tt NSIDE} \times {\rm \tt NSIDE}$ grids. Each pixel of a ${\rm \tt HEALPix}$ supersampled ${\rm \tt NSIDE}=4096$ map was filled according to the value at the corresponding RA and DEC in the reconstructed maps. The supersampled high ${\rm \tt NSIDE}$ maps were then degraded to ${\rm \tt NSIDE}=1024$. The true convergence maps from the simulations were directly binned from the convergence values at galaxy positions to ${\rm \tt NSIDE}=1024$. For all maps the same mask is applied, where pixels with no galaxies are masked.

Figure~\ref{fig:desmap} shows the mass map reconstructions from the SV shear data using the three different methods. An example simulation with truth and the three reconstructed maps is shown in figure~\ref{fig:simmap}. The ``tuning parameters'', $\sigma_{smooth} = 10.0 \rm{\ arcmin}$ for Kaiser-Squires and $\lambda = 3.0$ for {\sc Glimpse}, are tuned to maximise the Pearson correlation coefficient $r$ with the underlying truth when tested on simulations.

Using a suite of 10 simulations, in Sec.~\ref{sec:pearson} we calculate the Pearson correlation coefficient between the truth and the reconstruction with different methods as a test of the reconstruction's quality. In Sec.~\ref{sec:rms}, we calculate the root-mean-square error of the residuals between the truth and the reconstruction. In Sec.~\ref{sec:var} we calculate the variance of the 1-point distribution of the pixel values in the reconstruction and compare with the truth. In Sec.~\ref{sec:phase} and Sec.~\ref{sec:peaks} we quantify the quality of the reconstruction of the phase and peak statistics respectively, by comparing to the simulated truth. The final result presented in Sec.~\ref{sec:redmapper_results} compares the reconstruction from the DES SV shear data with foreground galaxy clusters from the redMaPPer catalogue (which are expected to trace non-linearities in the underlying density field). 

In this work we do not use correlation functions as a test of the map reconstruction. None of the mass mapping methods here are expected to reproduce the correct correlation functions or power spectra. It is simple to show this analytically with the Wiener filter, where despite the filter giving the MAP pixel values, the pixel variance, and therefore the power spectrum, is suppressed.

\begin{figure*}
\centering
\hspace*{-0.03in}
\includegraphics[width=1.04\textwidth]{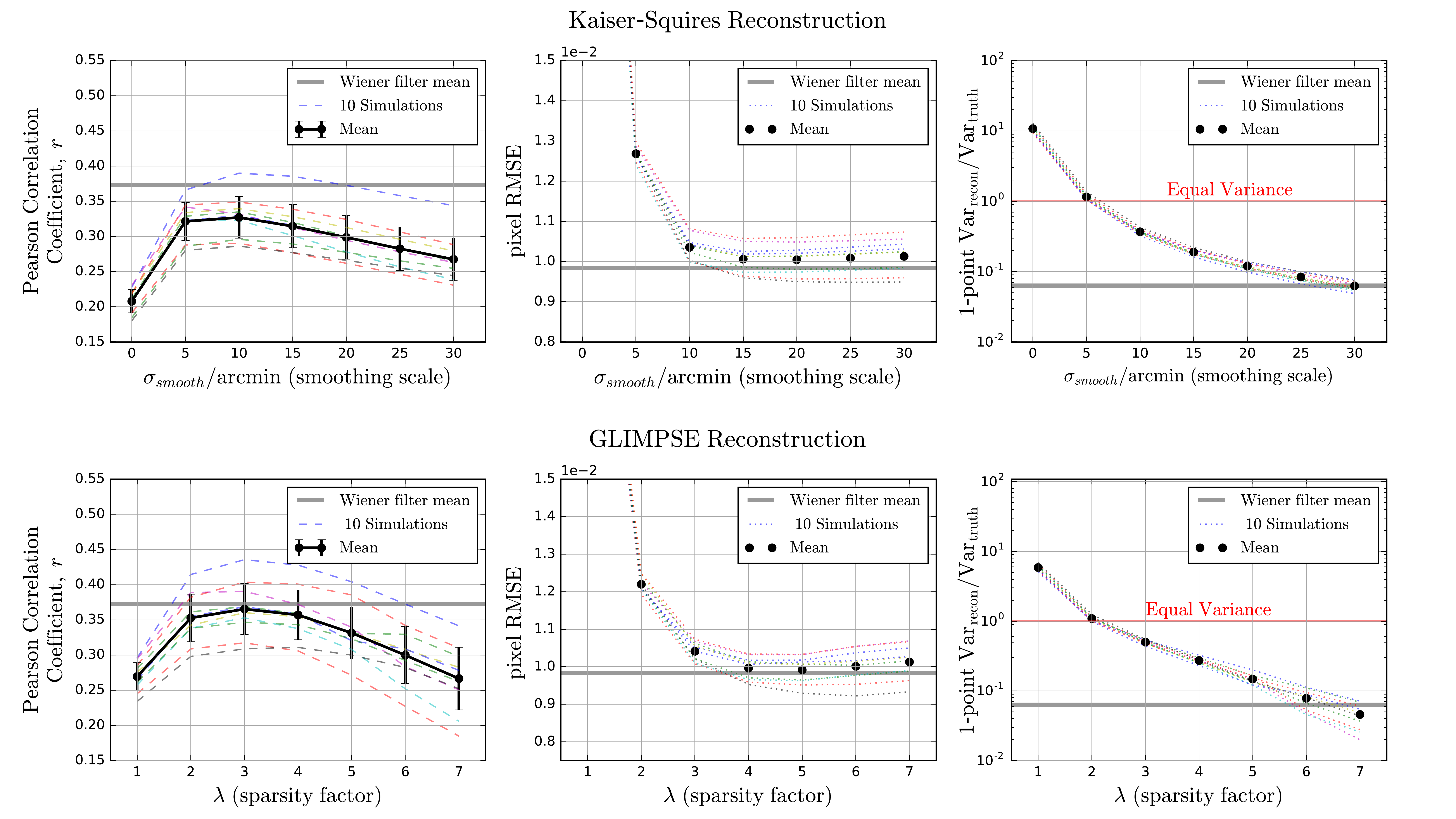}
\caption{\label{fig:pixels} Kaiser-Squires (top) and {\sc Glimpse} (bottom). Three different statistics comparing the true $\kappa$ map and the reconstruction with 10 simulations. \textit{Left panel}: The Pearson correlation coefficient, $r$ (equation~\ref{eq:pearson_r}). The errorbar on the mean is the standard deviation of the sample. The better the reconstruction, the higher the value of $r$. \textit{Middle panel}: The lower the pixel RMSE (equation~\ref{eq:rmse}), the better the reconstruction. \textit{Right panel}: Ratio of variances between the 1-point distribution of the pixels in the reconstruction and pixels in the true map (equation~\ref{eq:1pointvar}).}
\end{figure*} 

\subsection{Pixel Cross Correlation} \label{sec:pearson}

We quantify the correlation between the true convergence from simulation and the reconstructed convergence of the simulated catalogue using the Pearson correlation coefficient. As with other metrics of success for mass map reconstruction, this can be used to tune the sparsity $\lambda$ parameter and the smoothing scale for Kaiser-Squires.

The Pearson correlation coefficient, $r$, between the pixels' true convergence, $\kappa^{truth}$, and the reconstruction, $\kappa^{\rm{recon}}$, is given by

\begin{equation} \label{eq:pearson_r}
r =\frac{\sum^n_{i=1}(\kappa_i^{truth} - \bar{\kappa}^{truth})(\kappa_i^{\rm{recon}} - \bar{\kappa}^{\rm{recon}})}{\sqrt{\sum^n_{i=1}(\kappa_i^{truth} - \bar{\kappa}^{truth})^2} \sqrt{\sum^n_{i=1}(\kappa_i^{\rm{recon}} - \bar{\kappa}^{\rm{recon}})^2}} \ ,
\end{equation}

\noindent where the summations are over all pixels $i$ in the map and $\bar{\kappa}$ is the mean convergence in the map. 

In the left panels of figure~\ref{fig:pixels}, the Pearson $r$ value from 10 simulations is plotted for varying tuning parameters. Almost all of the simulations and also their mean have a maximal Pearson $r$ value at $\sigma_{smooth} = 10.0 \rm{\ arcmin}$ for Kaiser-Squires and at $\lambda = 3.0$ for {\sc Glimpse}. 

Table~\ref{table:pearson} presents the mean value from the 10 simulations, where the tuning parameter is chosen to maximise $r$ when relevant. All methods show good correlation with the underlying true convergence. Both the Wiener filter and {\sc Glimpse} have the same highest value of $r = 0.37$, $12 $ per cent higher than Kaiser-Squires. 

Note that the Pearson correlation coefficient as presented in equation~\ref{eq:pearson_r} is invariant under a rescaling of the reconstruction. Despite the Wiener filter reconstruction having values closer to zero, the Wiener filter maps still have good correlation to the truth. This second aspect is addressed in Sec.~\ref{sec:var} and in the second column of table~\ref{table:pearson}.

\subsection{Pixel Residuals} \label{sec:rms}

The difference between the true convergence from simulation and the reconstruction in pixel $i$ is defined as

\begin{equation}
\Delta \kappa_i = \kappa_i^{truth} - \kappa_i^{\rm{recon}} \ .
\end{equation}

\noindent We define the root-mean-square error (RMSE) as

\begin{equation} \label{eq:rmse}
\rm{RMSE}(\kappa^{truth}, \kappa^{\rm{recon}}) = \sqrt{\frac{1}{n} \sum_{i=1}^n \Delta \kappa_i^2} \ 
\end{equation}

\noindent where $n$ is the number of pixels. 

A smaller value of RMSE for a given method implies a better reconstruction according to this metric. It is this RMSE that the Wiener filter attempts to minimise using a linear filter, as defined in equation~\ref{eq:wiener_rmse}, by using an assumed signal covariance $\langle \mathbf{\kappa} \mathbf{\kappa^\dagger} \rangle$ (see Sec.~\ref{sec:wiener_theory}).

The centre panel of figure~\ref{fig:pixels} shows that increasing the smoothing scale, $\sigma_{smooth}$, for Kaiser-Squires or the regularisation parameter, $\lambda$, for {\sc Glimpse} initially reduces the pixel RMSE, but increased filtering contributes little beyond $\sigma_{smooth} = 10.0 \rm{\ arcmin}$ for Kaiser-Squires or $\lambda = 3.0$ for {\sc Glimpse}. 

The smallest mean pixel RMSE is $1.0 \times 10 ^{-2}$ for Kaiser-Squires and $9.9 \times 10 ^{-3}$ for {\sc Glimpse}. The Wiener filter, whose smoothing is constrained by the prior on $C_\ell$ and which therefore cannot be tuned, has a pixel RMSE of $9.4 \times 10^{-3}$.

\begin{figure}
\includegraphics[width=0.47\textwidth]{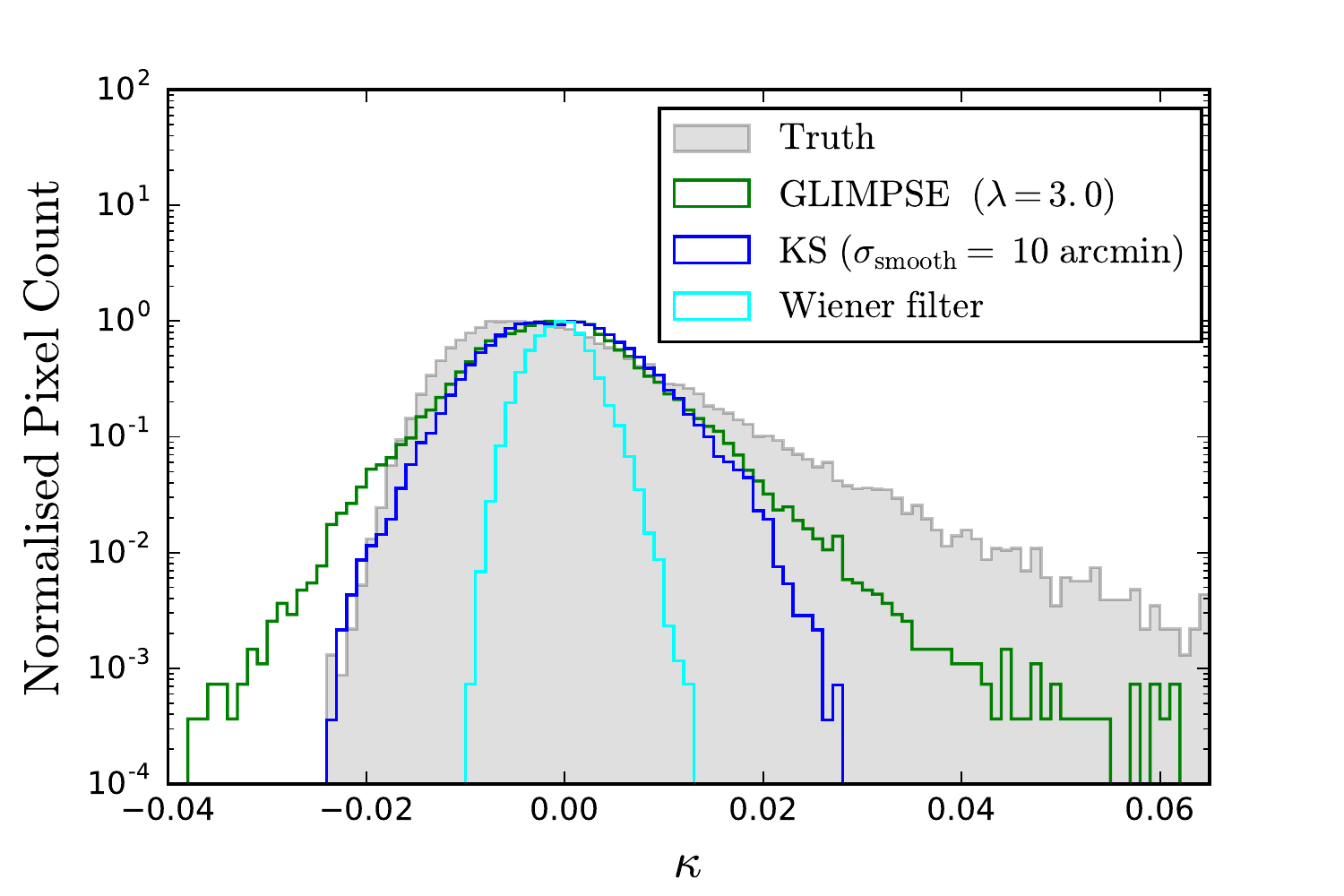}
\caption{\label{fig:kappa_hist} Pixel histograms (1-point distributions) for various map reconstructions from the simulated data shown in figure~\ref{fig:simmap}. The histograms are normalised such that the largest value of each is equal to one. The ratio of the variance between the reconstructions and the truth is presented in table~\ref{table:pearson}.}
\end{figure} 

\subsection{Pixel 1-Point Variance} \label{sec:var}

The 1-point distribution can be thought as a histogram of the pixel values. Figure~\ref{fig:kappa_hist} shows an example of such a histogram (derived from the simulated truth map and reconstructions of figure~\ref{fig:simmap}).

The mean of this distribution is unconstrained by weak lensing, due to an integration constant in equation~\ref{eq:kappa}. The variance of the 1-point distribution is increased compared to the underlying truth due to shape noise in the unsmoothed Kaiser-Squires reconstruction. A reconstruction method would aim to reduce the variance of the 1-point pixel distribution to match that of the underlying truth.

We define the estimate of the variance of the 1-point distributions of the truth or reconstructed $\kappa$ as

\begin{equation} \label{eq:1pointvar}
\begin{split}
\rm{Var}_{\rm{truth}} = &\frac{1}{n-1} \sum^n_{i=1}(\kappa_i^{\rm{truth}} - \bar{\kappa}^{\rm{truth}})^2 \\ 
\rm{Var}_{\rm{recon}} = &\frac{1}{n-1} \sum^n_{i=1}(\kappa_i^{\rm{recon}} - \bar{\kappa}^{\rm{recon}})^2 \ ,
\end{split}
\end{equation}

\noindent where the notation matches equation~\ref{eq:rmse}. The ratio of these variances is given by

\begin{equation} \label{eq:var_ratio}
\frac{\rm{Var}_{\rm{recon}}}{\rm{Var}_{truth}} = \frac{\sum^n_{i=1}(\kappa_i^{\rm{recon}} - \bar{\kappa}^{\rm{recon}})^2 }{\sum^n_{i=1}(\kappa_i^{truth} - \bar{\kappa}^{truth})^2 } \ ,
\end{equation}

\noindent The closer this value is to 1, the better the variance of the pixel distribution matches the truth. Using 10 simulations we can calculate this quantity for different reconstruction methods (and at different smoothing scales or $\lambda$ regularisation values where relevant). 

In figure~\ref{fig:pixels} the right panel shows the result of this test for {\sc Glimpse} and Kaiser-Squires. Both methods show a pixel distribution that has too high variance for insufficient filtering, and too low variance for over-filtering. For Kaiser-Squires, the ratio is closest to 1 at a smoothing scale of $\sigma_{smooth} = 5 \rm{\ arcmin}$. For {\sc Glimpse}, the ratio is closest to 1 at a sparsity regularisation value of $\lambda = 2$.

\begin{table}
\caption{The centre column gives the average Pearson correlation coefficient $r$ (equation~\ref{eq:pearson_r}) between $\kappa^{truth}$ and $\kappa^{\rm{recon}}$ from 10 simulations. The choices of $\sigma_{smooth} = 10 \rm{\ arcmin}$ and $\lambda = 3.0$ maximise the Pearson $r$ value. The right column gives the ratio of the pixel variance between $\kappa^{\rm{recon}}$ and $\kappa^{truth}$ (equation~\ref{eq:var_ratio}).} \label{table:pearson}
% \centering
\begin{tabular}{c c c}
\hline\hline 
 Method &  Pearson $r$  & Variance Ratio \\ [0.5ex] 
\hline
KS  ($\sigma_{smooth} = 10 \ \mathrm{arcmin}$) & 0.33  & 3.7 $\times 10^{-1}$  \\ Wiener filter   & 0.37 & 6.3 $ \times 10^{-2}$\\
{\sc Glimpse} ($\lambda = 3.0 $ )& 0.37 & 5.0 $\times 10^{-1}$ \\ [1ex] % [1ex] adds vertical space
\hline %inserts single line
\end{tabular}
\end{table}

Both of these reconstruction methods have a matching variance at a smoothing parameter value less than that which maximises the Pearson correlation coefficient $r$. If one chose this parameter to maximise the Pearson $r$ value, such that $\lambda = 3$ and $\sigma_{smooth} = 10 \rm{\ arcmin}$, a good reconstruction should also have the ratio of the variances as close to 1 as possible. 

The right column of table~\ref{table:pearson} gives the mean variance ratio from 10 simulations with the different methods. The choice of $\lambda = 3.0$ and $\sigma_{smooth} = 10 \rm{\ arcmin}$ are the tuning parameters that maximise the Pearson $r$ value for {\sc Glimpse} and Kaiser-Squires respectively. Though {\sc Glimpse} and the Wiener filter reconstructions both have the same Pearson $r$ value, the variance of the pixel values of the Wiener filter is much lower with respect to the underlying truth than is the case for {\sc Glimpse}. This can also be seen in the reconstructions of figure~\ref{fig:simmap}, where the Wiener filter pixel values are closer to zero than the simulated true convergence.

The histogram of figure~\ref{fig:kappa_hist} shows, for one single example, the distributions matching what the results of the second column of table~\ref{table:pearson} describe. {\sc Glimpse} outperforms the other methods at matching the variance of the underlying truth, however it still falls short. Also, all methods, including {\sc Glimpse}, have distributions which are symmetric, unlike the asymmetric, heavy-tailed distribution of the true $\kappa$ values. 

Though {\sc Glimpse} reconstructs maps with the 1-point distribution variance closest to the truth, it is also the only method to have convergence values dropping below the truth. These unphysical ``negative peaks'' can also be seen in the map reconstructions from data (figure~\ref{fig:desmap}) and from simulated catalogues (figure~\ref{fig:simmap}), and are likely to come from enforcement of sparsity for positive and negative wavelets equally. The physical motivation for {\sc Glimpse} comes from a density field of superimposed halos. Though there should be no negative halos, negative wavelets are included to map the underdense regions, clearly at the expense of producing these very negative regions.

\subsection{Phase reconstruction} \label{sec:phase}

The summation over all $m$ modes at each $\ell$ multipole in the angular power spectrum (equation~\ref{eq:cl}) loses all phase information; only the magnitudes are retained. This phase information corresponds to the spatial distribution of anisotropies. As the phases are dependent on the physical underlying structure, they contain information beyond what can be gained by 2-point statistics. Their retention is a well-motivated, desired property of a mass mapping reconstruction. 

Inspired by \cite{chapman_2012}, who use phases to test the reconstruction after foreground removal from simulated Epoch of Reonization 21-cm maps, we use the phase residual as a metric of success between our three methods.

The phase difference between the true map and the reconstruction is defined as

\begin{equation} \label{eq:phase_residual}
\begin{split}
\Delta \theta_{\ell m} &= \theta_{\ell m}^{truth} - \theta_{\ell m}^{\rm{recon}} \\
&= \mathrm{arg}\big(a^{truth}_{\ell m}\big) - \mathrm{arg}\big(a^{\rm{recon}}_{\ell m}\big) \ , \\
\ &\ \mathrm{where} \ \ \mathrm{arg}(z) = \mathrm{arctan}\Big(\frac{\mathrm{Im}(z)}{\mathrm{Re}(z)}\Big) \ .
\end{split}
\end{equation}

\noindent A small phase difference $\Delta \theta_{\ell m}$ between the truth and the reconstruction implies that the phase has been well reconstructed. For random variables drawn from a Gaussian distribution, this would correspond to a small standard deviation. Here, however, a Gaussian distribution would be an inappropriate choice as it assumes the data are defined on an unbounded Euclidean space.

The two dimensional data space of phase pairs, $ \{\theta_{\ell m}^{truth}, \theta_{\ell m}^{\rm{recon}} \}$, is a torus, $T^2$, and the projected data space of the phase difference, $\Delta \theta_{\ell m}$, is a circle, $S^1$. On a circle, the maximum entropy, least informative, distribution for specified mean and variance is the von Mises~\citep{JammalamadakaS.Rao2001Tics}, which in one dimension is given by

\begin{equation} \label{eq:phase_one}
Pr(\Delta \theta_{\ell m} | C, \mu) = \frac{1}{2  \pi  I_0 ( C )} \exp \big[ C \cos (\Delta \theta_{\ell m} - \mu)  \big] \ ,
\end{equation}

\noindent where $I_0$ is the modified Bessel function of order $0$, and $C$ is a concentration parameter. For $\mu = 0$, a large concentration parameter (analogous to $1/\sigma^2$) would correspond to a small dispersion in the phase reconstruction error. The aim is therefore to compare the inferred value of the concentration, $C$, between different mass mapping methods, with a larger value of $C$ implying a better phase reconstruction. 

By assuming that the error on the phase reconstruction is independent between phases, we can say that the phase differences, $\Delta \vv{\theta}$, are independent and identically distributed random variables, with a likelihood distribution given by

\begin{equation} \label{eq:phase_likelihood}
\begin{split}
Pr(\Delta \vv{\theta} | C, \mu) &= \prod_{\ell m} \frac{1}{2  \pi  I_0 ( C )}  \exp \big[  C \cos (\Delta \theta_{\ell m} - \mu)  \big]\\
& = \frac{1}{[2  \pi  I_0 ( C )]^n} \exp \big[ C \sum_{\ell m} \cos (\Delta \theta_{\ell m} - \mu)  \big] \ .
\end{split}
\end{equation}

\noindent As only the relative values of $C$ are needed to compare different mass mapping methods, the full posterior distribution is not required. Additionally, any reasonable prior distribution, $Pr(C)$, will be either flat or monotonically decreasing above zero, so the ranking of maps by the largest maximum likelihood value or maximum posterior value of $C$ will be identical. For the purposes of this comparison the simpler maximum likelihood estimate, $\hat{C}_{\rm{MLE}}$, will therefore do.

\begin{figure}
\hspace*{-0.1in}
\includegraphics[width=0.47\textwidth]{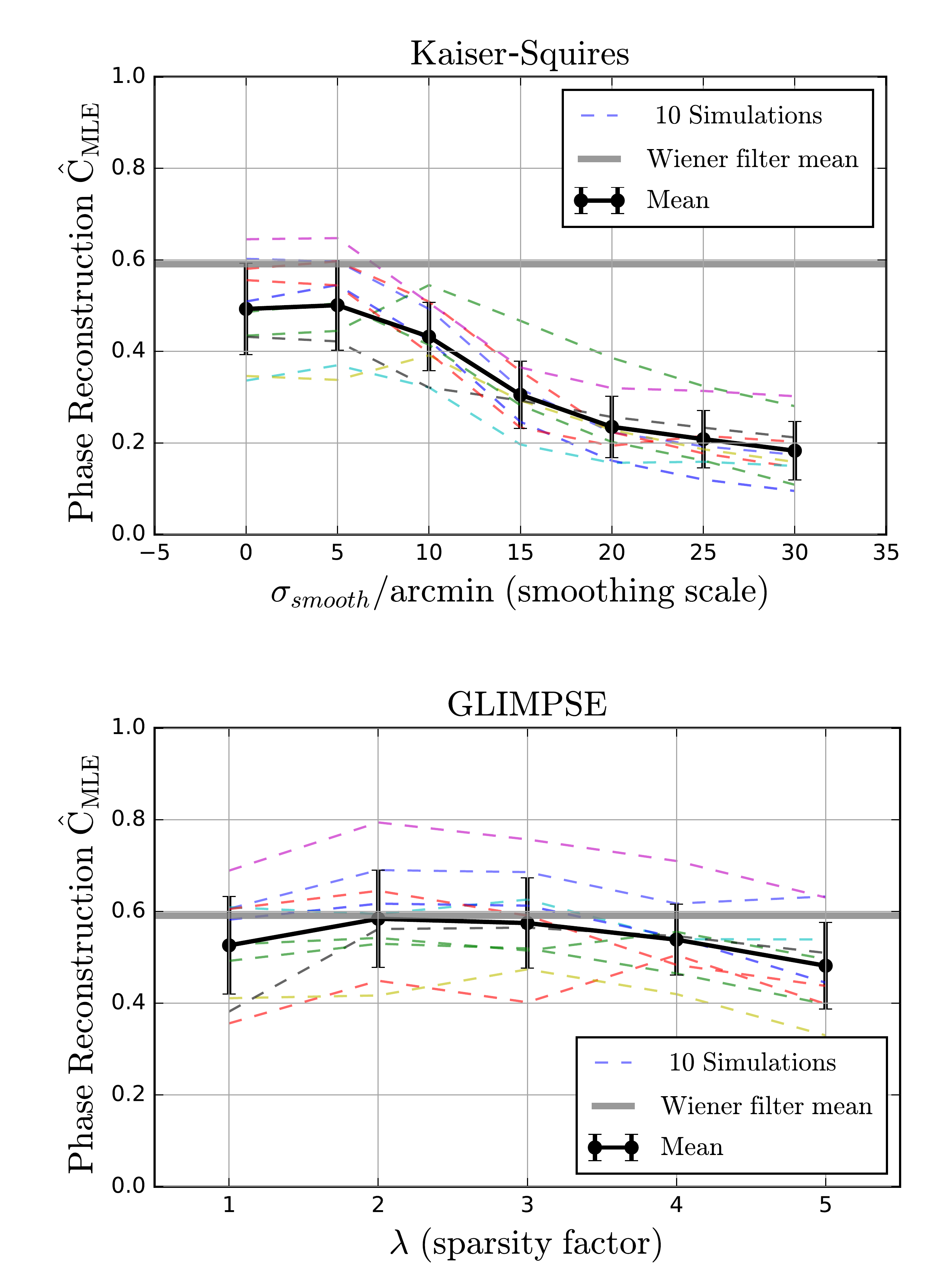}
\caption{\label{fig:ks_phase} The maximum likelihood value of the concentration of the phase residual distribution, $\hat{C}_{\rm{MLE}}$, as described by equation~\ref{eq:phase_likelihood}. The $\hat{C}_{\rm{MLE}}$ values are shown for 10 different simulations and with Kaiser-Squires (\textit{top panel}) at varying smoothing scale, $\sigma_{smooth}$, and {\sc Glimpse} (\textit{bottom panel}) at varying regularisation parameter $\lambda$. The phase reconstruction is best for $\sigma_{smooth} = 5 \rm{\ arcmin}$ and $\lambda = 3.0$ respectively.}
\end{figure} 

We calculate the maximum likelihood values of $\mu$ and $C$ by taking the spherical harmonic transform of our ${\rm \tt HEALPix}$ map to recover the $a_{\ell m}$ coefficients up to $\ell_{max} = 1024$, calculating the phase residual as defined by equation~\ref{eq:phase_residual} between the truth and the reconstruction for each coefficient, and then maximising the likelihood (equation~\ref{eq:phase_likelihood}). The maximisation is performed using the ${\rm \tt scipy}$ package BFGS algorithm~(\citealt{BFGS_1},~\citealt{BFGS_2},~\citealt{BFGS_3}), using $3$ random initialisation values to test for robustness.

Figure~\ref{fig:ks_phase} show the results for the phase reconstruction from 10 simulations using Kaiser-Squires and {\sc Glimpse} with varying tuning parameters. For Kaiser-Squires the mean phase reconstruction value, $\hat{C}_{\rm{MLE}}$, is maximised at $\sigma_{smooth}=5.0$ arcmin. For larger smoothing scales the phase reconstruction quality drops, as phase information is lost. For {\sc Glimpse} the mean phase reconstruction value, $\hat{C}_{\rm{MLE}}$, is maximised at $\lambda = 3.0$. The maximum value of $\hat{C}_{\rm{MLE}}$ is not particularly pronounced, and the $\hat{C}_{\rm{MLE}}$ values are quite stable over a range of $\lambda$.

\begin{table}
\caption{The mean over 10 simulations of the von Mises concentration maximum likelihood estimate, $\hat{C}_{\rm{MLE}}$, from phase residuals (equation~\ref{eq:phase_likelihood}).} \label{table:phases}
\centering
\begin{tabular}{c c}
\hline\hline 
 Method &  Phase reconstruction \\ \ & Concentration $\hat{C}_{\rm{MLE}}$  \\ [0.5ex] 
\hline
KS  ($\sigma_{smooth} = 5 \ \mathrm{arcmin}$) & 0.501  \\ Wiener filter   & 0.591 \\
{\sc Glimpse} ($\lambda = 3.0 $ ) & 0.584 \\ [1ex] % [1ex] adds vertical space
\hline %inserts single line
\end{tabular}
\end{table}

Table~\ref{table:phases} presents the mean values of $\hat{C}_{\rm{MLE}}$ with the best tuning parameters for the three map reconstruction methods. Both {\sc Glimpse} and the Wiener filter do much better than Kaiser-Squires for reconstructing the phases. Though the variance from these 10 different simulations is large, the Wiener filter does slightly better than {\sc Glimpse}, as can be seen in figure~\ref{fig:ks_phase}.

\subsection{Peak Statistics} \label{sec:peaks}

Peak statistics are a promising method for inferring cosmological parameters from data, as they access information beyond what can be inferred from 2-point correlation functions. Unlike higher order correlation functions, such as the bispectrum, peak statistics are inherently high signal-to-noise. They also probe the highly non-linear regions, where non-Gaussianity is greatest. The effect of masking is trivially taken into account by applying the identical mask to the suite of simulations used to construct a likelihood.

We cannot truly test which mass mapping method best constrains cosmology with the statistics of density peaks without fully deriving the posterior probability distributions of cosmological parameters. It is possible to test which method returns peaks which are distinguishable from noise and at which convergence values. Distinguishing a large number of peaks from noise at high values of $\kappa$ would mean the map is reconstructing the non-linear regions well.

For a given convergence map, we can define a function, $n(\kappa)$, that gives the number of peaks as a function of convergence. For a given mass reconstruction method we can compare the peaks in reconstructions from simulated data with the peaks in reconstructions from catalogues of ``randoms'', with shape noise but no weak lensing shear signal (equivalent to $\gamma = 0$ in equation~\ref{eq:shape_noise}). If a given map from data or from a simulated catalogue has the same $n(\kappa)$ as the random catalogues, then the mass mapping method used has been useless for peak statistics. On the other hand, if the map from data or simulation has a very different $n(\kappa)$ function to that from the reconstruction from the random catalogue, then the map reconstruction method has recovered ``true'', physical $\kappa$ peaks.

\begin{figure}
\centering
\hspace*{-0.1in}
\includegraphics[width=0.5\textwidth]{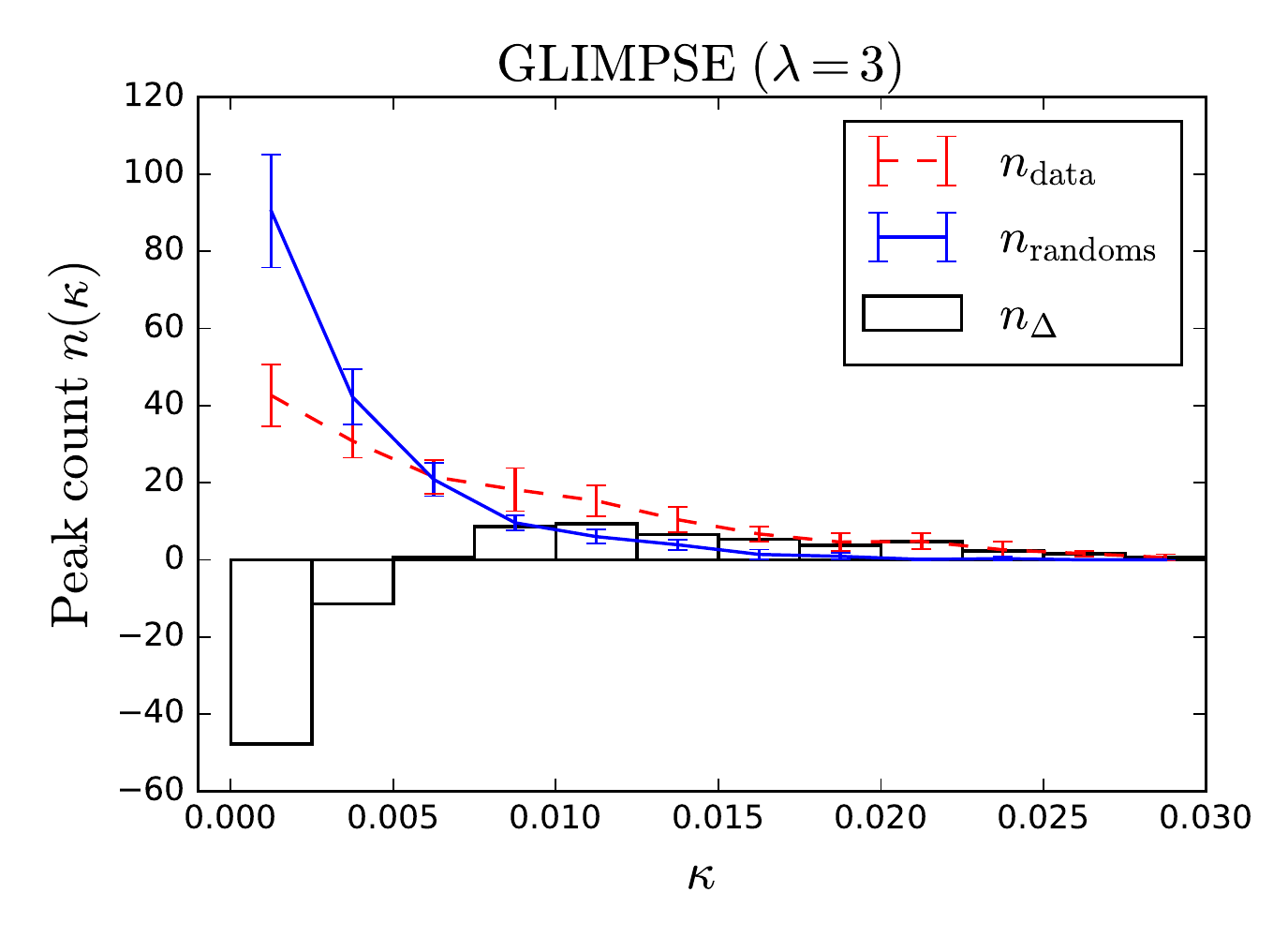}
\caption{\label{fig:peaks1} The mean $n_{\rm{data}}$, $n_{\rm{randoms}}$ and $n_\Delta$ functions with 10 simulated data catalogues and 10 random catalogues from {\sc Glimpse} ($\lambda = 3$) reconstructions. $n_\Delta$ is defined in equation~\ref{eq:peak_ndelta}. Errorbars are standard deviation sample estimates from the 10 simulations, and are consistent with Poissonian noise. Figure~\ref{fig:snr4} shows the signal-to-noise of $n_\Delta$ using the estimated Poissonian noise for different reconstruction methods and tuning parameters.}
\end{figure} 

\begin{figure*}
\centering
\hspace*{-0.01in}
\includegraphics[width=1.0\textwidth]{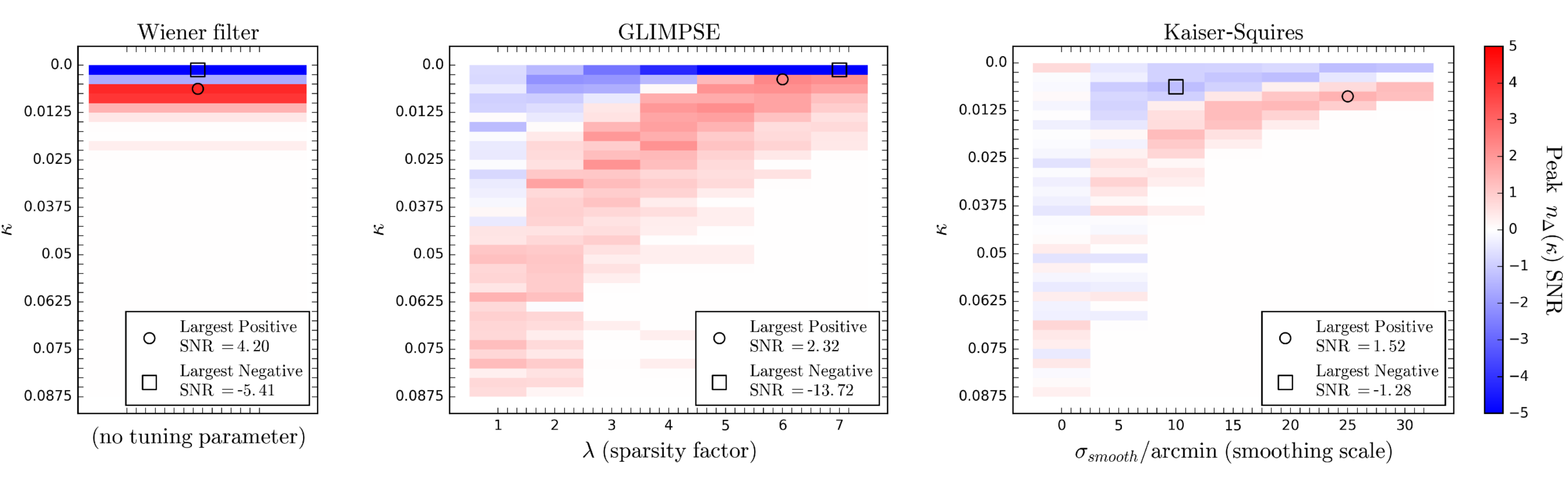}
\caption{\label{fig:snr4} The estimated signal-to-noise (SNR) of $n_\Delta (\kappa)$ (equation~\ref{eq:peak_ndelta}) using 10 simulated data catalogues and 10 random catalogues with the three different mass mapping methods. Kaiser-Squires and {\sc Glimpse} maps can be tuned by their respective parameters. The width of the left (Wiener filter) panel is purely nominal; it does not actually have a ``flat structure'', just no parameter to tune.}
\end{figure*} 

In the DES SV cosmology constraints from peak statistics,~\cite{des_peaks} use this difference as the data vector used to constrain cosmology,

\begin{equation} \label{eq:peak_ndelta}
n_\Delta(\kappa_i) = n_{\rm{data}}(\kappa_i) - n_{\rm{randoms}}(\kappa_i) \ .
\end{equation}

\noindent This function is far from zero at a given $\kappa$ if there is a large difference between the number of peaks counted in maps reconstructed (a) from data and (b) from random catalogues.

It is reasonable to believe that the number of peaks, $n(\kappa_i)$, in the $i^{th}$ bin, $\kappa_i$, is drawn from a Poisson distribution. The difference between two Poissonian random variables follows the Skellam distribution. Using this distribution, we expect the difference in the number peaks in maps from data and from random catalogues to have a mean given by

\begin{equation}
\mu_\Delta(\kappa_i) = \mu_{\rm{data}}(\kappa_i) - \mu_{\rm{randoms}}(\kappa_i) \ ,
\end{equation}

\noindent and a variance given by

\begin{equation} \label{eq:skellamvariance}
\sigma^2_{\Delta}(\kappa_i) = \mu_{\rm{data}}(\kappa_i) + \mu_{\rm{randoms}}(\kappa_i) .
\end{equation}

We can therefore define a peak signal-to-noise estimate

\begin{equation} \label{eq:peaksnr}
\rm{SNR}(\kappa_i) = \frac{\mu_\Delta(\kappa_i)}{\sqrt{\sigma^2_{\Delta}(\kappa_i) }} \  .
\end{equation}

Figure~\ref{fig:peaks1} shows $n_{\rm{data}}$, $n_{\rm{randoms}}$, and $n_\Delta$ from {\sc Glimpse} ($\lambda = 3$) from 10 simulations and 10 random catalogues. Here we define a peak as a local maxima in the {\tt HEALPIX} map. Across different methods and smoothing parameters, the predicted variance from equation~\ref{eq:skellamvariance} matches well with the estimated sample variance, verifying that the peak distribution is indeed Poissonian for a given $\kappa$.

Figure~\ref{fig:snr4} shows the peak signal-to-noise (SNR) estimates from 10 simulations and from 10 random catalogues as a function of $\kappa$ and smoothing scale, for Kaiser-Squires, or $\lambda$, for {\sc Glimpse}. As the peaks in the maps from data have higher convergence values than those from random catalogues, the $\rm{SNR}(\kappa)$ function is negative for low values of $\kappa$. 

In the figures, the {\sc Glimpse} reconstruction gives better signal-to-noise estimates on the peaks than does the Kaiser-Squires reconstruction. For Kaiser-Squires, the largest positive and negative signal-to-noise values are 1.52 and -1.28. For {\sc Glimpse}, the largest positive and negative are signal-to-noise values of 2.32 and -13.72. For the Wiener filter these values are 4.20 and -5.41.

The Wiener filter therefore has the highest signal-to-noise of the peak function $n_\Delta (\kappa)$, though the $\kappa$ values of these peaks are very low. As can be seen in the reconstruction from the SV data in figure~\ref{fig:desmap}, the pixel values of the Wiener filter are much closer to zero. This is reflected in the peak statistic signal-to-noise values. In the left panel of figure~\ref{fig:snr4}, the Wiener filter detects negligibly few peaks with $\kappa > 0.0125$, whereas {\sc Glimpse} detects peaks with positive signal-to-noise up to higher values of $\kappa$. It is at these high values where the non-Gaussian information due to non-linear structure formation can be probed.

\subsection{Foreground Clusters} \label{sec:redmapper_results}

Comparisons with foreground clusters of galaxies is an independent test of the mass map reconstructions, as it uses data (unlike our tests on simulations).

\begin{figure*}
\hspace*{-0.9in}
\centering
\includegraphics[width=1.25\textwidth]{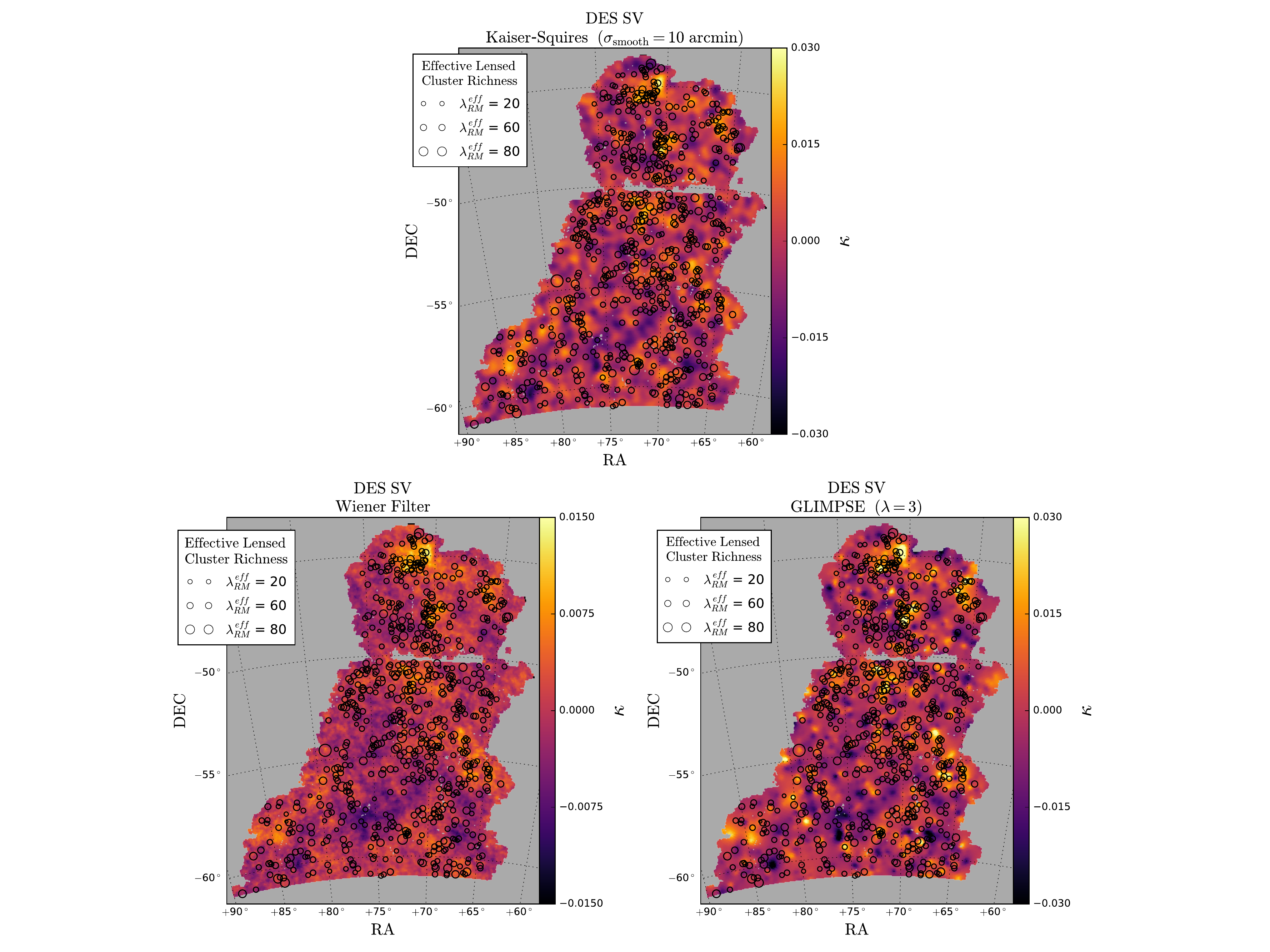}
\caption{\label{fig:clustersmap} The mass map reconstruction from DES SV shear data with the three different methods, as presented in figure~\ref{fig:desmap}), with the locations of redMaPPer clusters overlaid. The size of the cluster marker is the effective richness of the cluster, as defined in equation~\ref{eq:effective_richness}. \textbf{Note} that the colour scale for the Wiener filter is less than that for the other reconstructions, as the pixel values are closer to zero.}
\end{figure*}

In figure~\ref{fig:clustersmap} the redMaPPer clusters described in Sec.~\ref{sec:redmapper} are overlaid on the DES SV $\kappa$ map reconstructions shown in figure~\ref{fig:desmap}. The maps show good spatial correlation between the locations of the clusters and the $\kappa$ peaks in the map. 

The size of a cluster marker is the effective lensed cluster richness $\lambda^{eff}_{RM}$, rather than the redMaPPer cluster richness. This concept is adapted from the definition of $\kappa_{g}$ presented in~\cite{chang_sv_map}. For a given cluster, this is defined as

\begin{equation} \label{eq:effective_richness}
\lambda^{eff}_{RM} = \frac{p(z) \omega(z)}{a(z)} \times \lambda_{RM} \times \frac{\langle \lambda_{RM} \rangle}{\langle \lambda^{eff}_{RM} \rangle} \  ,
\end{equation}

\noindent where $z$ is the redshift of the cluster, $p(z)$ is the lensing efficiency at the location of the cluster (see figure~\ref{fig:pz}), and $\omega(z)$ is the comoving distance to the cluster (so that the first term matches the integrand of equation~\ref{eq:Q2}). The final term normalises the mean, where $\langle \lambda_{RM} \rangle$ is the average richness over all galaxy clusters. The effective lensed cluster richness gives the richness as ``seen'' by the lensing effect, where clusters at the peak of the lensing efficiency should contribute more to the $\kappa$ map. We therefore calculate the correlation between $\lambda^{eff}_{RM}$ for each cluster and the reconstructed $\kappa$ value at the cluster centre. 

This method does not take into account multiple clusters overlapping in a given line of sight. In figure~\ref{fig:clustersmap}, many small clusters overlap on large peaks in the reconstructed $\kappa$ map. The naive one-to-one correspondence between cluster and $\kappa$ would mistake this for an excess of $\kappa$ in the reconstruction. However, all methods will suffer equally from this assumption. A more thorough treatment of this overlapping effect is left for future work.

\begin{table}
\caption{The Pearson correlation coefficient value, $r$,  between effective richness, $\lambda^{eff}_{RM}$, of the foreground redMaPPer clusters and the reconstructed convergence map at the location of each galaxy cluster.} \label{table:redmapper}
\centering
\begin{tabular}{c c}
\hline\hline 
 Method & redMaPPer Cluster $\lambda^{eff}_{RM}$  \\ \ & Pearson $r$  \\ [0.5ex] 
\hline
KS  ($\sigma_{smooth} = 10 \ \mathrm{arcmin}$) & 0.116  \\ Wiener filter   & 0.129 \\
{\sc Glimpse} ($\lambda = 3.0 $ ) & 0.152 \\ [1ex] % [1ex] adds vertical space
\hline %inserts single line
\end{tabular}
\end{table}

Table~\ref{table:redmapper} presents the Pearson correlation coefficient $r$ between the $\lambda^{eff}_{RM}$ value of each cluster and the $\kappa^{\rm{recon}}$ value at the corresponding pixel. The tuning parameters for Kaiser-Squires and {\sc Glimpse} are chosen to maximise the Pearson correlation coefficient $r$ between the reconstruction and the truth from simulations (see Sec.~\ref{sec:pearson}). 

Though both {\sc Glimpse} and the Wiener filter take into account the noise and the mask in the data, and therefore do better than Kaiser-Squires, the {\sc Glimpse} reconstructions show higher correlation with the effective richness of the foreground clusters than do the Wiener filter reconstructions. This is no surprise, as {\sc Glimpse} is expected to do better at reconstructing non-Gaussian $\kappa$, which would correspond to the non-linear matter structures in which clusters of galaxies form.

\section{Conclusions}

In this work we have presented convergence map reconstructions using the public DES SV shear data with three different methods: Kaiser-Squires, Wiener filter, and {\sc Glimpse}. Kaiser-Squires is a simple inversion from shear to convergence, whereas the Wiener filter and {\sc Glimpse} use prior knowledge about the true convergence to help regularise the reconstruction and to reduce the effects of noise and missing data. The Wiener filter is a Bayesian MAP estimate if the signal and noise are Gaussian and the respective covariance matrices are known. The {\sc Glimpse} method enforces a sparsity-promoting $l_1$ norm in a wavelet space where the wavelets represent positive, isotropic, quasi-spherical objects well. {\sc Glimpse} is therefore expected to do well at reconstructing non-linear structures. The Wiener filter and {\sc Glimpse} therefore aim to reconstruct different regimes: the linear and non-linear density field. 

The three methods were applied to realistic simulations of the DES SV shear data, for which an underlying true convergence is known. Using these simulations we are also able to tune the Kaiser-Squires smoothing scale, $\sigma_{smooth}$, and the {\sc Glimpse} sparsity regularisation parameter, $\lambda$.

With these simulations we measure the Pearson correlation coefficient, $r$, between the truth and the reconstruction with different methods. Compared to the Kaiser-Squires reconstructions we find a $12 $ per cent improvement in Pearson correlation with both the Wiener filter and {\sc Glimpse}. The tuning parameters of $\sigma_{smooth} = 10 \rm{\ arcmin}$ for Kaiser-Squires and $\lambda = 3$ for {\sc Glimpse} maximise the Pearson correlation. We also measure the variance of the 1-point distribution of the reconstructed convergence. The Wiener filter suppresses the variance to 6.3 {\rm \ per\ cent} of the truth, Kaiser-Squires to 37 {\rm \ per\ cent} and {\sc Glimpse} to 50 {\rm \ per\ cent} of the truth. The tunable parameters here were those which maximised the Pearson correlation with the truth.

A large motivation for creating these maps is to reconstruct the convergence while still retaining the non-Gaussian information (which cannot be accessed with 2-point statistics such as the power spectrum). As such, we test the reconstruction of the harmonic phases, which is averaged out in the power spectrum, and the signal-to-noise of a peak statistic data vector, which is a popular probe of non-Gaussian information. The phase residuals between the truth and the reconstruction have the highest von Mises concentration with the Wiener filter ($\hat{C}_{\rm{MLE}} = 0.591$), with the {\sc Glimpse} reconstruction performing comparably ($\hat{C}_{\rm{MLE}} = 0.584$). Both methods outperformed the Kaiser-Squires reconstruction ($\hat{C}_{\rm{MLE}} = 0.501$). 

With realistic data vectors for peak statistics generated from simulations, the maximum signal-to-noise value was increased by a factor of 3.5 for the Wiener filter and by a factor of 9 for {\sc Glimpse}, compared to Kaiser-Squires. The signal-to-noise of the peak statistic data vector ($n_\Delta (\kappa)$) is shown in figure~\ref{fig:snr4}, where {\sc Glimpse} has significant signal-to-noise with high convergence peaks, where non-linearities in the underlying density field are highest. We predict these high value peaks are most useful for constraining cosmology beyond Gaussianity. In order to constrain cosmology with these different reconstruction methods, realistic simulations with different cosmological parameters or models must be used and the same reconstruction method should be applied to the simulations and data. As seen from our results, different reconstruction methods can produce convergence maps with different properties.

Finally, we switched from using simulations to instead using real observations (DES SV data). Here we measured the correlation between the reconstructed maps and the effective richness of the foreground redMaPPer clusters (this is the cluster richness as ``seen'' by the lensing effect). Table~\ref{table:redmapper} shows the results. Compared with Kaiser-Squires, the Wiener filter shows a $18 $ per cent increase and {\sc Glimpse} shows a $32 $ per cent increase in correlation. This demonstrates with independent, cosmological data the ability of the methods to reconstruct non-linear structures.

The metrics we have used for comparing the three reconstruction methods are generic, and they have been inspired by recent applications of weak lensing mass maps to cosmological studies (e.g.~\citealt{changbias},~\citealt{des_peaks}). These metrics may not be optimal for evaluating every application of mass maps. Future studies can compare the efficiency of the three and other methods in end-to-end analyses; for example, with the estimation of cosmological parameters or identification of galaxy clusters.

Applying the Wiener filter and {\sc Glimpse} methods to the DES Year 1 (Y1) shear catalogue would require extensions of the methods to account for the curved sky at large angular scales. The Y1 data covers $\approx 1500 \rm{\ deg}^2$ and contains $\approx 34,800,000$ galaxies, so is a large increase in data volume from DES SV. This modification has already been done with an extension of Kaiser-Squires to the sphere by~\cite{y1_mass_map} for the Y1 DES data. These extensions would also be useful for the upcoming $\approx 5000 \rm{\ deg}^2$ DES Y3 shear catalogue.

Of future interest would be to use the Wiener filter or {\sc Glimpse} convergence maps for scientific results, as we have shown that they reconstruct the convergence better than Kaiser-Squires according to many different metrics. 

\vspace{14mm} 

We have made our map reconstructions (as shown in figure~\ref{fig:desmap}) available at \url{des.ncsa.illinois.edu/releases/sva1}.

\vspace{7mm} 

\section*{Acknowledgements}

NJ, FBA and J-LS acknowledge support from the European Community through the DEDALE grant (contract no. 665044) within the H2020 Framework Program of the European Commission. OL acknowledges support from a European Research Council Advanced Grant FP7/291329 and support from the UK Science and Technology Research Council (STFC) Grant No. ST/M001334/1. FBA also acknowledges the support of the Royal Society for a University Research Fellowship.

DJJ acknowledges the support of the National Science Foundation, award AST-1440254.

We are grateful for the extraordinary contributions of our CTIO colleagues and the DECam Construction, Commissioning and Science Verification teams in achieving the excellent instrument and telescope conditions that have made this work possible.  The success of this project also relies critically on the expertise and dedication of the DES Data Management group.

Funding for the DES Projects has been provided by the U.S. Department of Energy, the U.S. National Science Foundation, the Ministry of Science and Education of Spain, the Science and Technology Facilities Council of the United Kingdom, the Higher Education Funding Council for England, the National Center for Supercomputing Applications at the University of Illinois at Urbana-Champaign, the Kavli Institute of Cosmological Physics at the University of Chicago, the Center for Cosmology and Astro-Particle Physics at the Ohio State University, the Mitchell Institute for Fundamental Physics and Astronomy at Texas A\&M University, Financiadora de Estudos e Projetos, Funda{\c c}{\~a}o Carlos Chagas Filho de Amparo {\`a} Pesquisa do Estado do Rio de Janeiro, Conselho Nacional de Desenvolvimento Cient{\'i}fico e Tecnol{\'o}gico and the Minist{\'e}rio da Ci{\^e}ncia, Tecnologia e Inova{\c c}{\~a}o, the Deutsche Forschungsgemeinschaft and the Collaborating Institutions in the Dark Energy Survey. 

The Collaborating Institutions are Argonne National Laboratory, the University of California at Santa Cruz, the University of Cambridge, Centro de Investigaciones Energ{\'e}ticas, Medioambientales y Tecnol{\'o}gicas-Madrid, the University of Chicago, University College London, the DES-Brazil Consortium, the University of Edinburgh, the Eidgen{\"o}ssische Technische Hochschule (ETH) Z{\"u}rich, Fermi National Accelerator Laboratory, the University of Illinois at Urbana-Champaign, the Institut de Ci{\`e}ncies de l'Espai (IEEC/CSIC), the Institut de F{\'i}sica d'Altes Energies, Lawrence Berkeley National Laboratory, the Ludwig-Maximilians Universit{\"a}t M{\"u}nchen and the associated Excellence Cluster Universe, the University of Michigan, the National Optical Astronomy Observatory, the University of Nottingham, The Ohio State University, the University of Pennsylvania, the University of Portsmouth, SLAC National Accelerator Laboratory, Stanford University, the University of Sussex, Texas A\&M University, and the OzDES Membership Consortium.

Based in part on observations at Cerro Tololo Inter-American Observatory, National Optical Astronomy Observatory, which is operated by the Association of Universities for Research in Astronomy (AURA) under a cooperative agreement with the National Science Foundation.

The DES data management system is supported by the National Science Foundation under Grant Numbers AST-1138766 and AST-1536171. The DES participants from Spanish institutions are partially supported by MINECO under grants AYA2015-71825, ESP2015-66861, FPA2015-68048, SEV-2016-0588, SEV-2016-0597, and MDM-2015-0509, some of which include ERDF funds from the European Union. IFAE is partially funded by the CERCA program of the Generalitat de Catalunya. Research leading to these results has received funding from the European Research Council under the European Union's Seventh Framework Program (FP7/2007-2013) including ERC grant agreements 240672, 291329, and 306478. We  acknowledge support from the Australian Research Council Centre of Excellence for All-sky Astrophysics (CAASTRO), through project number CE110001020, and the Brazilian Instituto Nacional de Ci\^encia e Tecnologia (INCT) e-Universe (CNPq grant 465376/2014-2).

This manuscript has been authored by Fermi Research Alliance, LLC under Contract No. DE-AC02-07CH11359 with the U.S. Department of Energy, Office of Science, Office of High Energy Physics. The United States Government retains and the publisher, by accepting the article for publication, acknowledges that the United States Government retains a non-exclusive, paid-up, irrevocable, world-wide license to publish or reproduce the published form of this manuscript, or allow others to do so, for United States Government purposes.

We use the visualization software package {\rm \tt SkyMapper}\footnote{\url{https://github.com/pmelchior/skymapper}} for the map figures.

%%%%%%%%%%%%%%%%%%%%%%%%%%%%%%%%%%%%%%%%%%%%%%%%%%



\bibliographystyle{mnras}
\bibliography{bibliog} % if your bibtex file is called example.bib

%%%%%%%%%%%%%%%%%%%%%%%%%%%%%%%%%%%%%%%%%%%%%%%%%%

%%%%%%%%%%%%%%%%% APPENDICES %%%%%%%%%%%%%%%%%%%%%

% \appendix

\appendix
\section{Author Affiliations}
\label{sec:affiliations}
{\small
$^{1}$ Department of Physics \& Astronomy, University College London, Gower Street, London, WC1E 6BT, UK\\
$^{2}$ Department of Physics and Electronics, Rhodes University, PO Box 94, Grahamstown, 6140, South Africa\\
$^{3}$ McWilliams Center for Cosmology, Department of Physics, Carnegie Mellon University, Pittsburgh, PA 15213, USA\\
$^{4}$ Laboratoire AIM, UMR CEA-CNRS-Paris 7, Irfu, SAp/SEDI, Service d'Astrophysique, CEA Saclay, 91191 Gif-sur-Yvette Cedex, France\\
$^{5}$ Kavli Institute for Cosmological Physics, University of Chicago, Chicago, IL 60637, USA\\
$^{6}$ Department of Physics and Astronomy, University of Pennsylvania, Philadelphia, PA 19104, USA\\
$^{7}$ Department of Physics, ETH Zurich, Wolfgang-Pauli-Strasse 16, CH-8093 Zurich, Switzerland\\
$^{8}$ Max Planck Institute for Extraterrestrial Physics, Giessenbachstrasse, 85748 Garching, Germany\\
$^{9}$ Universit\"ats-Sternwarte, Fakult\"at f\"ur Physik, Ludwig-Maximilians Universit\"at M\"unchen, Scheinerstr. 1, 81679 M\"unchen, Germany\\
$^{10}$ Argonne National Laboratory, 9700 South Cass Avenue, Lemont, IL 60439, USA\\
$^{11}$ Cerro Tololo Inter-American Observatory, National Optical Astronomy Observatory, Casilla 603, La Serena, Chile\\
$^{12}$ Fermi National Accelerator Laboratory, P. O. Box 500, Batavia, IL 60510, USA\\
$^{13}$ Institute of Cosmology \& Gravitation, University of Portsmouth, Portsmouth, PO1 3FX, UK\\
$^{14}$ Instituto de Fisica Teorica UAM/CSIC, Universidad Autonoma de Madrid, 28049 Madrid, Spain\\
$^{15}$ CNRS, UMR 7095, Institut d'Astrophysique de Paris, F-75014, Paris, France\\
$^{16}$ Sorbonne Universit\'es, UPMC Univ Paris 06, UMR 7095, Institut d'Astrophysique de Paris, F-75014, Paris, France\\
$^{17}$ Laborat\'orio Interinstitucional de e-Astronomia - LIneA, Rua Gal. Jos\'e Cristino 77, Rio de Janeiro, RJ - 20921-400, Brazil\\
$^{18}$ Observat\'orio Nacional, Rua Gal. Jos\'e Cristino 77, Rio de Janeiro, RJ - 20921-400, Brazil\\
$^{19}$ Department of Astronomy, University of Illinois at Urbana-Champaign, 1002 W. Green Street, Urbana, IL 61801, USA\\
$^{20}$ National Center for Supercomputing Applications, 1205 West Clark St., Urbana, IL 61801, USA\\
$^{21}$ Institut de F\'{\i}sica d'Altes Energies (IFAE), The Barcelona Institute of Science and Technology, Campus UAB, 08193 Bellaterra (Barcelona) Spain\\
$^{22}$ Institut d'Estudis Espacials de Catalunya (IEEC), 08193 Barcelona, Spain\\
$^{23}$ Institute of Space Sciences (ICE, CSIC),  Campus UAB, Carrer de Can Magrans, s/n,  08193 Barcelona, Spain\\
$^{24}$ Kavli Institute for Particle Astrophysics \& Cosmology, P. O. Box 2450, Stanford University, Stanford, CA 94305, USA\\
$^{25}$ Centro de Investigaciones Energ\'eticas, Medioambientales y Tecnol\'ogicas (CIEMAT), Madrid, Spain\\
$^{26}$ Department of Physics, IIT Hyderabad, Kandi, Telangana 502285, India\\
$^{27}$ Department of Astronomy/Steward Observatory, 933 North Cherry Avenue, Tucson, AZ 85721-0065, USA\\
$^{28}$ Jet Propulsion Laboratory, California Institute of Technology, 4800 Oak Grove Dr., Pasadena, CA 91109, USA\\
$^{29}$ Department of Astronomy, University of Michigan, Ann Arbor, MI 48109, USA\\
$^{30}$ Department of Physics, University of Michigan, Ann Arbor, MI 48109, USA\\
$^{31}$ SLAC National Accelerator Laboratory, Menlo Park, CA 94025, USA\\
$^{32}$ Center for Cosmology and Astro-Particle Physics, The Ohio State University, Columbus, OH 43210, USA\\
$^{33}$ Department of Physics, The Ohio State University, Columbus, OH 43210, USA\\
$^{34}$ Harvard-Smithsonian Center for Astrophysics, Cambridge, MA 02138, USA\\
$^{35}$ Australian Astronomical Observatory, North Ryde, NSW 2113, Australia\\
$^{36}$ Departamento de F\'isica Matem\'atica, Instituto de F\'isica, Universidade de S\~ao Paulo, CP 66318, S\~ao Paulo, SP, 05314-970, Brazil\\
$^{37}$ Department of Astrophysical Sciences, Princeton University, Peyton Hall, Princeton, NJ 08544, USA\\
$^{38}$ Instituci\'o Catalana de Recerca i Estudis Avan\c{c}ats, E-08010 Barcelona, Spain\\
$^{39}$ School of Physics and Astronomy, University of Southampton,  Southampton, SO17 1BJ, UK\\
$^{40}$ Department of Physics, Brandeis University, Waltham, MA 02453, USA\\
$^{41}$ Instituto de F\'isica Gleb Wataghin, Universidade Estadual de Campinas, 13083-859, Campinas, SP, Brazil\\
$^{42}$ Computer Science and Mathematics Division, Oak Ridge National Laboratory, Oak Ridge, TN 37831\\
}

\section{Indicator Function} \label{sec:indicator}

We define the indicator function $i_{\mathcal{C}}$ (as used in equation~\ref{eq:optimisation}) of a set $\mathcal{C}$ as
\begin{equation}
i_{\mathcal{C}}(x) = \begin{cases}
	0 & \mbox{if } x \in \mathcal{C}\\
	+ \infty & \mbox{otherwise}  \ .
\end{cases}
\end{equation}

%%%%%%%%%%%%%%%%%%%%%%%%%%%%%%%%%%%%%%%%%%%%%%%%%%

% Don't change these lines
\bsp	% typesetting comment
\label{lastpage}
\end{document}